\documentclass[11pt,letterpaper]{amsart}

\usepackage[foot]{amsaddr}
\usepackage{amsmath}
\usepackage{amsthm}
\usepackage{amssymb}
\usepackage{mathrsfs}
\usepackage{url}
\usepackage{hyperref}
\usepackage{todonotes}
\usepackage{mathtools}
\usepackage[left=1in,right=1in,top=1in,bottom=1in]{geometry}

\newtheorem{Thm}{Theorem}
\newtheorem{Lem}[Thm]{Lemma}

\newtheorem{Prob}[Thm]{Problem}
\newtheorem{Cor}[Thm]{Corollary}
\newtheorem{Conj}[Thm]{Conjecture}
\theoremstyle{remark}

\theoremstyle{definition}

\newenvironment{Proof}{\begin{proof}}{\end{proof}}

\newcommand{\bra}{\ensuremath{\langle}}
\newcommand{\ket}{\ensuremath{\rangle}}
\newcommand{\Z}{\ensuremath{\mathbb{Z}}}
\newcommand{\E}{\ensuremath{\mathbb{E}}}
\newcommand{\F}{\ensuremath{\mathbb{F}}}
\newcommand{\K}{\ensuremath{\mathrm{K}}}
\newcommand{\Qm}{\ensuremath{\mathrm{Q}}}
\newcommand{\C}{\ensuremath{\mathbb{C}}}
\newcommand{\Q}{\ensuremath{\mathbb{Q}}}
\newcommand{\cC}{\ensuremath{\mathcal{C}}}
\newcommand{\cF}{\ensuremath{\mathcal{F}}}
\newcommand{\cG}{\ensuremath{\mathcal{G}}}
\newcommand{\cH}{\ensuremath{\mathcal{H}}}
\newcommand{\sP}{\ensuremath{\mathscr{P}}}
\newcommand{\rk}{\ensuremath{\operatorname{rk}}}
\newcommand{\supp}{\ensuremath{\operatorname{supp}}}
\newcommand{\bad}{\ensuremath{\operatorname{bad}}}
\newcommand{\good}{\ensuremath{\operatorname{good}}}
\newcommand{\poly}{\ensuremath{\operatorname{poly}}}
\newcommand{\Ra}{\ensuremath{\mathrm{R}}}
\newcommand{\Rb}{\ensuremath{\mathrm{\underline{R}}}}
\newcommand{\Ri}{\ensuremath{\mathscr{R}}}
\newcommand{\bO}{\ensuremath{\mathcal{O}}}
\newcommand{\btO}{\ensuremath{\mathcal{\tilde O}}}
\newcommand{\dcup}{\,\dot{\cup}\,}

\newcommand{\YL}{\stackrel{\raisebox{1pt}{\tiny\text{Yates}}}{\longleftarrow}}
\newcommand{\iv}[1]{[\![{#1}]\!]}
\DeclareMathSymbol{\downsym}{\mathopen}{symbols}{"23}
\newcommand{\down}{\ensuremath\operatorname{\downsym}}

\begin{document}


\title[]{%
The Asymptotic Rank Conjecture and\\{}
the Set Cover Conjecture\\{}
are not Both True}

\author{Andreas Bj\"orklund}
\address{IT University of Copenhagen}
\email{anbjo@itu.dk}

\author{Petteri Kaski}
\address{Aalto University, Department of Computer Science}
\email{petteri.kaski@aalto.fi}

\begin{abstract}
Strassen's asymptotic rank conjecture [{\em Progr.~Math.}~120~(1994)] 
claims a strong submultiplicative upper bound on the rank of a three-tensor 
obtained as an iterated Kronecker product of a constant-size base tensor. 
The conjecture, if true, most notably would put square matrix multiplication 
in quadratic time. We note here that some more-or-less unexpected algorithmic 
results in the area of exponential-time algorithms would also follow. 
Specifically, we study the so-called set cover conjecture, which states 
that for any $\epsilon>0$ there exists a positive integer constant $k$ 
such that no algorithm solves the $k$-Set Cover problem 
in worst-case time $\bO((2-\epsilon)^n|\mathcal F|\operatorname{poly}(n))$. 
The $k$-Set Cover problem asks, given as input 
an $n$-element universe $U$, a family $\mathcal F$ of 
size-at-most-$k$ subsets of $U$, and a positive integer $t$, whether there 
is a subfamily of at most $t$ sets in $\mathcal F$ whose union is $U$.
The conjecture was formulated by Cygan \emph{et al.}~in the monograph 
{\em Parameterized Algorithms} [Springer,~2015] but was implicit as 
a hypothesis already in 
Cygan \emph{et al.}~[CCC~2012, {\em ACM~Trans.~Algorithms}~2016], 
there conjectured to follow from the Strong Exponential Time Hypothesis.
We prove that if the asymptotic rank conjecture is true, then the 
set cover conjecture is false. 
Using a reduction by Krauthgamer and Trabelsi [STACS~2019], 
in this scenario we would also get an $\bO((2-\delta)^n)$-time 
randomized algorithm for some constant $\delta>0$ for another 
well-studied problem for which no such algorithm is known, namely 
that of deciding whether a given $n$-vertex directed graph has 
a Hamiltonian cycle.
\end{abstract}

\maketitle

\thispagestyle{empty}


\setcounter{page}{0}

\clearpage

\section{Introduction}

\noindent
This paper has two protagonists, the \textsc{Set Cover} problem 
and a certain constant-size tensor.
\begin{Prob}[\textsc{($k$-)Set Cover}]
\label{prob:set-cover}
Given as input 
(i) an $n$-element universe $U$, 
(ii) a family $\cF=\{S_1,S_2,\ldots,S_m\}$
of subsets $S_j\subseteq U$ with $|S_j|\leq k$ for $1\leq j\leq m$, and 
(iii) a positive integer $t$, 
decide whether there exist $1\leq j_1< j_2< \cdots< j_{t'}\leq m$ for some $t'\leq t$
such that $\bigcup_{\ell=1}^{t'} S_{j_\ell}=U$. 
\end{Prob}

Since the introduction of dynamic programming by Bellman
in the 1950s~\cite{Bellman1957}, it has been known that 
the \textsc{Set Cover} problem 
can be solved by algorithms that run in worst-case 
time~$\bO^*(2^n)$; cf.~Fomin, Kratsch, and Woeginger~\cite{FominKW2004}.%
\footnote{%
The $\bO^*(\cdot)$ notation suppresses a factor polynomial in the input size.}{}
In the last two decades, there has been renewed interest in the study
of algorithms that run in exponential time 
(e.g.~Fomin and Kratsch~\cite{FominK2010}), 
with the \textsc{Set Cover} problem attracting both attention as well as
notoriety in terms of absence of discoveries of algorithms with faster 
running time. This has led to the hypothesis or even the conjecture that 
no faster algorithms exist, made implicitly by 
Cygan~{\em et al.}~\cite{CyganDLMNOPSW2016} 
who conjectured that the Strong Exponential Time Hypothesis
of Impagliazzo and Paturi~\cite{ImpagliazzoP2001} implies the following 
hardness assumption for \textsc{Set Cover}, and explicitly recorded in
Cygan {\em et al.}~\cite[Conjecture~14.36]{CyganFKLMPPS15} as well as by
Krauthgamer and Trabelsi~\cite{KrauthgamerT2019}:

\begin{Conj}[Set cover conjecture; 
Cygan~{\em et al.}~\cite{CyganFKLMPPS15};
Krauthgamer and Trabelsi~\cite{KrauthgamerT2019}]
\label{conj:set-cover}
For every fixed $\epsilon>0$ there exists a positive integer $k$ such that 
no algorithm (even randomized) solves 
the $k$-\textsc{Set Cover} problem in time $\bO^*(2^{(1-\epsilon)n})$.
\end{Conj}

In this paper, we connect the study of the set cover conjecture
into the study of multilinear algebra and the study of asymptotic tensor 
rank of tensors. From an algorithmic standpoint, the latter field was 
initiated by Strassen with his 1969 breakthrough result~\cite{Strassen1969} 
on the sub-cubic time complexity of matrix multiplication, which subsequently 
developed into a theory of multilinear algebraic complexity and the study
of asymptotic rank of tensors 
(e.g.~B\"urgisser {\em et~al.}~\cite{BurgisserCS1997} and 
Landsberg~\cite{Landsberg2012}). 
Our interest, however, is neither on tensors that represent 
matrix multiplication, nor is it directly motivated by such tensors. 
Rather, our second protagonist is the following $2\times 2\times 2$ tensor:
\begin{equation}
\label{eq:three-way-partitioning}
P=\left[\begin{array}{@{\,}c@{\ }|@{\ }c@{\,}}
\begin{array}{@{\,}c@{\,}c@{\,}}
0&1\\
1&0
\end{array} &
\begin{array}{@{\,}c@{\,}c@{\,}}
1&0\\
0&0
\end{array}
\end{array}\right]\,.
\end{equation}
This {\em three-way partitioning} tensor $P$ in its $n$\textsuperscript{th}
Kronecker power $P^{\otimes n}$ of shape $2^n\times 2^n\times 2^n$
indicates all three-tuples of sets
$X,Y,Z\subseteq [n]$ that satisfy $X\dcup Y\dcup Z=[n]$.
The tensor $P$ is also a tensor that is known to
satisfy (over any field $\F$) the conditions and the conclusion of 
the asymptotic rank conjecture for tensors made by Strassen in 
1994~\cite{Strassen1994} 
(see also~\cite{Strassen1987,Strassen1988,Strassen1991}), 
with recent interest in the conjecture in the algebraic geometry community 
(e.g.~\cite{ChristandlVZ2021,ConnerGLV2022,ConnerGLVW2021,Landsberg2019}):%
\footnote{We postpone precise 
definitions of tensor rank as well as concise and tight tensors 
to Section~\ref{sect:preliminaries}.}{}

\begin{Conj}[Asymptotic rank conjecture; Strassen~\cite{Strassen1994}]
\label{conj:asymptotic-rank}
Let\/ $\F$ be a field and let $Q\in\F^{c\times c\times c}$ be a tensor that
is concise and tight. Then, for all positive integers $n$, the tensor rank
of $Q^{\otimes n}$ is at most $c^{n+o(n)}$. 
\end{Conj}

\noindent
The asymptotic rank conjecture remains open for the vast majority of 
tensors, including the $4\times 4\times 4$ tensor that represents the
multiplication of two $2\times 2$ matrices---the conclusion of the conjecture
for this tensor would immediately imply $\omega=2$ for the exponent $\omega$
of square matrix multiplication (over all fields that share the
characteristic of $\F$); e.g.~\cite{BurgisserCS1997}.

In this paper, we identify a specific $7\times 7\times 7$ subtensor $Q$ 
of $P^{\otimes 3}$ such that the conclusion of the asymptotic rank conjecture 
(over any finite field or the field of complex numbers)
for this tensor~$Q$ falsifies the set cover conjecture. In particular, 
Conjecture~\ref{conj:set-cover} and Conjecture~\ref{conj:asymptotic-rank}
cannot both be true.

\subsection{Main result and corollaries}

Our main result is the following theorem. 

\begin{Thm}[Main; The set cover conjecture fails under the asymptotic rank conjecture]
\label{thm:main}
Let\/ $\F$ be either a finite field or the field of complex numbers. 
If the asymptotic rank conjecture is true over\/~$\F$, then there exist
constants $\kappa,\epsilon>0$ and a randomized algorithm that solves 
the $k$-\textsc{Set Cover} problem for $1\leq k\leq \kappa n$
in time $\bO((2-\epsilon)^n)$ with high probability. 
\end{Thm}

Before proceeding to give an overview of techniques underlying 
Theorem~\ref{thm:main}, let us record an immediate corollary via 
existing reductions in the literature; in particular, a reduction of 
Krauthgamer and Trabelsi~\cite{KrauthgamerT2019} gives 
the following corollary for the directed Hamiltonian cycle problem.

\begin{Cor}[Directed Hamiltonicity under the asymptotic rank conjecture]
Let\/ $\F$ be either a finite field or the field of complex numbers. 
If the asymptotic rank conjecture is true over\/~$\F$, then there exists
a constant $\epsilon>0$ and a randomized algorithm that given an $n$-vertex 
directed graph as input decides whether it has a Hamiltonian cycle 
in $\bO((2-\epsilon)^n)$ time. 
\end{Cor}

The key algorithmic tool that enables our main result is a randomized 
algorithm for the following {\em three-way partitioning} problem: given as 
input three families $\cF,\cG,\cH\subseteq 2^{[n]}$, decide whether there 
exist $(X,Y,Z)\in\cF\times\cG\times\cH$ such that $X\dcup Y\dcup Z=[n]$.
For a positive integer $k$, we say that $\cF,\cG,\cH$ are $k$-{\em bounded}
if all sets in $\cF,\cG,\cH$ have size at most $k$.
Our main theorem follows from the lemma below by routine combinatorial
reductions, which we postpone to~Sect.~\ref{sect:covering-and-partitioning}.

\begin{Lem}[Main; Three-way partitioning under the asymptotic rank conjecture]
\label{lem:main-three-way}
Let\/ $\F$ be either a finite field or the field of complex numbers. 
If the asymptotic rank conjecture is true over\/~$\F$, then there
exist constants $\delta,\tau>0$ as well as a randomized algorithm that solves 
the three-way partitioning problem with high probability 
in $\bO((2-\delta)^n)$ time for
$(\frac{1}{3}+\tau)n$-bounded families of subsets.
\end{Lem}

\subsection{Overview of techniques}

The algorithm design in Lemma~\ref{lem:main-three-way} stems from a 
combination of elementary and known observations, which we now proceed to
outline; we refer to Sects.~\ref{sect:preliminaries} and \ref{sect:three-way}
for notation and a detailed proof of Lemma~\ref{lem:main-three-way}.

\medskip
\noindent
{\em 1.~The multilinear framework and Yates's algorithm.}
First, the three-way partitioning
problem is equivalent to asking whether the $2^n\times 2^n\times 2^n$ 
tensor $T=P^{\otimes n}$ restricted to $\cF\times\cG\times\cH$ contains at
least one $1$-entry. Or, what is the same, writing $f,g,h$ for $2^n$-length 
zero-one-valued indicator vectors of the set families $\cF,\cG,\cH$, 
respectively, whether
the trilinear form $T[f,g,h]=\sum_{x,y,z\in [2^n]} T_{x,y,z}f_xg_yh_z$ 
is positive. Or, what is the same, but viewed as a bilinear operator 
on the vectors $g$ and $h$ defined by $T$, combined with the 
transpose of the vector $f$ by 
$f^\top T[g,h]=\sum_{x\in[2^n]} f_x \sum_{y,z\in [2^n]}T_{x,y,z}g_yh_z$.
The bilinear operator $(g,h)\mapsto T[g,h]$ is%
\footnote{Up to complementation with respect to $[n]$; 
cf.~Sect.~\ref{sect:fast-subset-convolution} for a precise statement.}{} 
the {\em subset convolution} operator, which is known to admit evaluation in 
$\btO(2^n)$ time,%
\footnote{The $\btO(\cdot)$ notation suppresses a multiplicative factor 
polylogarithmic in the parameter.}{}
or ``fast subset convolution''~\cite{BjorklundHKK2007}
(see also~\cite{Brand2022}),
by the principle of inclusion and exclusion, restriction by size, 
and Yates's algorithm~\cite{Yates1937}.

\medskip
\noindent
{\em 2.~Fast subset convolution via a border decomposition.}
Second, the three-way partitioning tensor $P$ is known to have border 
rank $2$ (cf.~Sect.~\ref{sect:ranks}), which enables 
direct trilinear algorithms with running time $\btO(2^n)$ by using the
border decomposition in Yates's algorithm~\cite{Yates1937}.
This in particular enables an apparently novel way to derive fast subset
convolution (cf.~Sect.~\ref{sect:fast-subset-convolution}), which however
still remains restricted to $\btO(2^n)$ time. Furthermore and in particular
due to the tensor $P^{\otimes n}$ having border rank $2^n$ since it 
contains the $2^n\times 2^n$ identity matrix as a matrix slice.

\medskip
\noindent
{\em 3.~Breaking the algorithm for asymptotic rank decrease.}
Third, we proceed to study the subtensors of $P^{\otimes n}$ to 
reduce the rank---that is, to enable faster algorithms 
via Yates's algorithm---while yet maintaining control on the size of the
support (that is, the number of $1$-entries in the tensor). This 
study of partial tensors and the support of a tensor
stems from earlier works of Sch\"onhage~\cite{Schonhage1981} and 
Cohn and Umans~\cite{CohnU2013} in the context of fast matrix multiplication,
which were further quantified into a ``breaking and randomizing'' framework
by Karppa and Kaski~\cite{KarppaK2019} for Boolean matrix multiplication,
utilizing the fact that matrix-multiplication tensors have support-transitive
symmetry. This notion of support-transitivity however does not  
apply to the three-way disjointness tensor, which necessitates more
intricate and non-uniform randomization. 
We apply the breaking and randomizing framework 
to the tensor $P^{\otimes c}$ for a positive integer constant $c$, 
starting with the elementary observation that the tensor has a support of 
size $3^c$ and border rank~$2^c$; thus, assuming we break the tensor by 
introducing $h$ holes 
(that is, we turn $h$ of the $1$-entries in the tensor into $0$-entries), 
and witness an {\em asymptotic} rank decrease of one, then, assuming 
that we can sufficiently randomize the locations of the $h$ holes, we can 
potentially salvage by independent random repetitions a correct 
sub-$2^n$-time randomized algorithm from the broken tensor 
when $\frac{3^c}{3^c-h}\cdot (2^c-1)<2^c$. 
This intuition suggests considering $c=h=3$ in particular
since $\frac{27}{24}\cdot 7<8$. 

\medskip
\noindent
{\em 4.~Realizing the rank decrease under the asymptotic rank conjecture.}
Fourth, we identify a subtensor $Q$ obtainable by breaking $P^{\otimes 3}$ by
introducing $3$ holes so that $Q$ becomes a 
$7\times 7\times 7$ tensor that is both tight and concise. 
Thus, under the asymptotic rank
conjecture, we then have that the rank of $Q^{\otimes d}$ is at most
$7^{\frac{1001}{1000}d}$ for a positive integer constant $d$. 

\medskip
\noindent
{\em 5.~Randomization to salvage a correct algorithm from the broken algorithm.}
Fifth, we design a randomization strategy that randomly permutes and trims
the given input $\cF,\cG,\cH$, with the effect of adequately randomizing
the location of the holes in the tensor $P^{\otimes p}\otimes Q^{\otimes dq}$ 
for nonnegative integers $p$ and $q$ such that $n=p+3dq$. This has probability
sufficienty close to $(24/27)^{dq}$ of detecting any fixed three-way partition
in the input when there exists one, at the cost of essentially
$\bO((|\cF|+|\cG|+|\cH|+2^p7^{\frac{1001}{1000}dq})\poly(n))$ time. 
By making essentially $(27/24)^{dq}\poly(n)$ independent random repetitions
of this algorithm with $q=\lfloor\frac{n}{1000d}\rfloor$, we obtain an 
algorithm that for $(\frac{1}{3}+\frac{1}{1000})n$-bounded inputs 
runs in overall 
$\bO((2-\frac{1}{100000})^n)$ time and succeeds with probability at least
$1-\exp(\Omega(n))$. Here we stress that we have optimized neither the 
analysis nor the construction of the tensor $Q$ from a running-time standpoint.

\subsection{Related work}

We now proceed with a review of select further related work around the two
conjectures and in the area of parameterized and exact exponential algorithms.

\medskip
\noindent
{\em Tensors and tensor rank.}
Since Strassen's breakthrough on matrix multiplication~\cite{Strassen1969},
numerous variants of the notion of tensor rank have been introduced to 
assist studying the rank of matrix multiplication tensors.
These include border rank~\cite{BiniCRL1979}, 
asymptotic rank~\cite{Gartenberg1985}, support rank~\cite{CohnU2013}, and
probabilistic rank~\cite{KarppaK2019}. The asymptotic rank 
conjecture~\cite[Conjecture~5.3]{Strassen1994} follows a sequence of works 
by Strassen on the notion of degeneration and the asymptotic spectrum of 
tensors~\cite{Strassen1987,Strassen1988,Strassen1991,Strassen1994}. 
Strassen's laser method and its improvements in particular underlie 
many advances in fast matrix multiplication
\cite{AlmanVW2021,CoppersmithW1990,LeGall2014,Stothers2010,VassilevskaWilliams2012}. 
The asymptotic rank conjecture has received recent interest in
the algebraic geometry community
(e.g.~\cite{ChristandlVZ2021,ConnerGLV2022,ConnerGLVW2021,Landsberg2019}).
In very recent work, Alman, Turok, Yu, and Zhang~\cite{AlmanTYZ2023} explore
tensor ranks and the complexity of dynamic programming.

\medskip
\noindent
{\em The low-rank method in algorithm design.} 
There is an extensive literature in the field of parameterized and 
exact exponential algorithms on using rank and rank-related parameters
of matrices as tools for algorithm design; we refer to 
Nederlof~\cite{Nederlof2020b} for a more extensive survey. 
Among key examples of this approach are the homomorphism-basis 
subgraph-counting technique of Curticapean, Dell, and 
Marx~\cite{CurticapeanDM2017}, the Hamiltonicity algorithm of Cygan,
Kratsch, and Nederlof via bases of perfect matchings~\cite{CyganKN2018},
the matrix-rank based counting techniques of Curticapean, Lindzey, and
Nederlof~\cite{CurticapeanLN2018} as well as 
Jansen and Nederlof~\cite{JansenN2019}, the representative set technique
of Fomin, Lokshtanov, and Saurabh~\cite{FominLS2014}, the
representative function technique of Lokshtanov, Saurabh, and 
Zehavi~\cite{LokshtanovSZ2021}, and the rank-based representative techniques
of Bodlaender, Cygan, Kratsch, and Nederlof~\cite{BodlaenderCKN2013}.
Specifically as regards low-rank techniques combined with 
a low-rank hypothesis for matrix multiplication tensors ($\omega=2$), 
Nederlof~\cite{Nederlof2020}
presents an $\bO(1.9999^n)$-time algorithm for the bipartite traveling 
salesperson problem subject to the hypothesis. 

\medskip
\noindent
{\em Set cover and set partitioning.}
A general instance $(U,\mathcal F,t)$ of the \textsc{Set Cover} problem 
over $n=|U|$ elements from $m=|\mathcal F|$ subsets can be solved 
in $\bO(2^nmn)$ time by dynamic programming using $\bO(2^n)$ space, 
as noted by Fomin~\emph{et al.}~\cite{FominK2010}. 
Karp~\cite{Karp82} observed that the principle of 
inclusion and exclusion can be used in place of dynamic programming in 
some cases to obtain the same running time but requiring only polynomial space.
Bj\"orklund and Husfeldt used Karp's technique for counting perfect matchings 
in a graph and other set partitioning problems~\cite{BjorklundH08}. 
Bj\"orklund~\emph{et al.}~\cite{BjorklundHK2009} observed that 
a set cover can also be computed in $\bO(n2^n+mn)$ time and space using 
Yates's algorithm to implement the zeta and M\"obius transforms 
on the subset lattice.
When the subsets in $\mathcal F$ are of size at most $k$, and we seek 
a partitioning of the universe $U$, faster algorithms exist. 
Koivisto~\cite{Koivisto2009} showed that the set partitions in this case 
can be counted in $\bO((2-\epsilon_k)^n)$ time, with $\epsilon_k>0$ 
a constant depending only on $k$. Also for the case of sets of size 
exactly $k$, Bj\"orklund~\cite{Bjorklund2010} gave 
a faster $\btO(2^{n(k-2)/k})$-time algorithm for the detection of 
a set partitioning. Another parameterization considers the solution 
size $t$. For a set cover consisting of $t=\sigma n$ sets, 
Nederlof~\cite{Nederlof16} gave 
an $\bO^*(2^{(1-\Omega(\sigma^4))n})$-time algorithm.
The graph coloring problem is a key variant of the set cover problem 
where the family consists of all the independent sets of the input graph.
Bj\"orklund~\emph{et al.}~\cite{BjorklundHK2009} gave an $\bO^*(2^n)$-time 
algorithm for deciding if a $k$-coloring exists in an $n$-vertex graph. 
Recently, Zamir~\cite{Zamir23} gave an $\bO((2-\epsilon_k)^n)$-time 
algorithm for $k$-coloring for some constant $\epsilon_k>0$ in regular graphs. 

\medskip
\noindent
{\em Directed Hamiltonicity.}
Since the discovery of an $\bO(1.657^n)$-time algorithm for 
Hamiltonicity detection in \emph{undirected} $n$-vertex graphs 
by Bj\"orklund~\cite{Bjorklund2014},
it has been open whether an algorithm running in time $\bO((2-\epsilon)^n)$ 
for some constant $\epsilon>0$ exists also for directed graphs.
This was also posed as Open Problem~6.20 by Husfeldt at a Dagstuhl 
Seminar in 2013~\cite{Husfeldt2013}.
There is some partial support for the existence of such an algorithm. 
Cygan, Kratsch, and Nederlof~\cite{CyganKN2018} showed that for bipartite 
directed graphs, there is an $\bO(1.888^n)$-time algorithm using 
a matrix-rank technique. A faster algorithm running in time 
$\btO(3^{n/2})$ for bipartite directed graphs, using different methods, 
was discovered by Bj\"orklund, Kaski, and Koutis~\cite{BjorklundKK2017}. 
Another direction considers modular counting of the Hamiltonian cycles. 
Bj\"orklund and Husfeldt~\cite{BjorklundH2013} showed that computing 
the parity of the number of Hamiltonian cycles can be done 
in $\bO(1.619^n)$ time. That result was subsequently generalized to 
address the detection problem when the Hamiltonian cycles in the graph 
are not too many. The best result we are aware of shows that if the graph 
has less than $d^n$ Hamiltonian cycles for \emph{any} constant $d$, 
then Hamiltonicity can be decided in $\bO((2-\epsilon_d)^n)$ time for 
some constant $\epsilon_d>0$~\cite{BjorklundKK2017}. Still, for the general 
graph case no $\bO((2-\epsilon)^n)$-time algorithm is known for 
the problem for any constant $\epsilon>0$.

\subsection{Organization of the paper}

The rest of this paper is organized as follows. 
Section~\ref{sect:preliminaries} presents notational, algebraic, and 
algorithmic preliminaries. Section~\ref{sect:three-way} proves our 
main technical result, Lemma~\ref{lem:main-three-way}, for finite fields.
Section~\ref{sect:covering-and-partitioning} proves Theorem~\ref{thm:main}
from Lemma~\ref{lem:main-three-way} by combinatorial reductions.
Appendix~\ref{sect:complex} proves Lemma~\ref{lem:main-three-way}
for the field of complex numbers.

\section{Preliminaries}

\label{sect:preliminaries}

We choose to work with tensors and matrices in coordinates to enable direct
algorithmic use as well as accessibility to communities with less background 
in algebra or geometry. All material in this section is well known, apart
possibly from the derivation of fast subset convolution using Yates's
algorithm directly from the border decomposition 
(cf.~Sect.~\ref{sect:fast-subset-convolution}); we are not aware of
earlier published work with this derivation.
For a geometric development of tensors, 
cf.~Landsberg~\cite{Landsberg2012,Landsberg2019}. 
For background in algebra and algebraic complexity theory, 
cf.~Lang~\cite{Lang2002} and 
B\"urgisser, Clausen, and Shokrollahi~\cite{BurgisserCS1997}. 

For a nonnegative integer $m$, let us write $[m]=\{0,1,\ldots,m-1\}$
and $2^{[m]}$ for the set of all subsets of $[m]$. 
We work with Iverson's bracket notation; for a logical proposition $\sP$, we
write $\iv{\sP}=1$ if $\sP$ is true and $\iv{\sP}=0$ if $\sP$ is false. 
For two sets $X$ and $Y$, we write $X\dcup Y$ for the union of $X$ and $Y$
and to highlight that $X$ and $Y$ are disjoint sets.
Let us write $\Z$ for the set of integers. 

The asymptotic notation $\bO^*(\cdot)$ suppresses a multiplicative factor 
polynomial in the input size, whereas the notation $\btO(\cdot)$ suppresses
a multiplicative factor polylogarithmic in the parameter. 

\subsection{Matrices, tensors, and products}

Let $\Ri$ be a commutative ring.
For a matrix $A\in\Ri^{m\times n}$ of {\em shape} $m\times n$, we write 
$A_{i,j}\in\Ri$ for the entry of $A$ at row $i\in [m]$ and column $j\in [n]$. 
All of our tensors will have order three. 
For a tensor $T\in\Ri^{m\times n\times p}$ of {\em shape} $m\times n\times p$, 
we write $T_{i,j,k}\in\Ri$ for the entry of $T$ at level $i\in [m]$, 
row $j\in [n]$, and column $k\in [p]$. A matrix or tensor is {\em zero} if
all of its entries are equal to zero; otherwise the matrix or tensor is 
{\em nonzero}. A column (row) {\em vector} is a matrix with one 
column (row). For a matrix $A\in\Ri^{m\times n}$, we write 
$A^\top\in\Ri^{n\times m}$ for the {\em transpose} of $A$, defined for 
all $i\in [m]$ and $j\in [n]$ by $(A^\top)_{j,i}=A_{i,j}$.

\medskip
\noindent
{\em Matrix product.}
For matrices $A\in\Ri^{m\times n}$ and $B\in \Ri^{n\times p}$,
the {\em product} $AB\in \Ri^{m\times p}$ is defined for all 
$i\in[m]$ and $j\in [p]$ by $(AB)_{i,j}=\sum_{k\in [n]}A_{i,k}B_{k,j}$.

\medskip
\noindent
{\em Kronecker product.}
For matrices $A\in\Ri^{m\times n}$ and $B\in \Ri^{u\times v}$,
the {\em Kronecker product} $A\otimes B\in\Ri^{mu\times nv}$ is defined 
for all $i\in [m]$, $j\in [n]$, $k\in[u]$, and $\ell\in[v]$ by
$(A\otimes B)_{iu+k,jv+\ell}=A_{i,j}B_{k,\ell}$. 
For tensors $S\in\Ri^{m\times n\times p}$ and $T\in \Ri^{u\times v\times w}$, 
the {\em Kronecker product} $S\otimes T\in\Ri^{mu\times nv\times pw}$ is defined 
for all $i\in [m]$, $j\in [n]$, $k\in[p]$, $a\in[u]$, $b\in [v]$, and 
$c\in [w]$ by $(S\otimes T)_{iu+a,jv+b,kw+c}=S_{i,j,k}T_{a,b,c}$. 
For a matrix or tensor $Z$ and a positive integer $p$, we write
$Z^{\otimes p}$ for the Kronecker product of $p$ copies of $Z$.
For example, $Z^{\otimes 3}=Z\otimes Z\otimes Z$.

\medskip
\noindent
{\em Hadamard product.}
For matrices or tensors $Z$ and $Z'$ of the same shape, 
their {\em Hadamard product} $Z\odot Z'$ is a matrix or tensor of the same shape
obtained by multiplying the entries of $Z$ and $Z'$. 

\medskip
\noindent
{\em Kruskal product~\cite{Kruskal1977}.}
For matrices $A\in\Ri^{m\times r}$, $B\in \Ri^{n\times r}$, and 
$C\in\Ri^{p\times r}$, the {\em Kruskal product} of $A,B,C$ is the
tensor $T\in\Ri^{m\times n\times p}$ defined for all $i\in [m]$, $j\in [n]$,
and $k\in [p]$ by $T_{i,j,k}=\sum_{\ell\in[r]}A_{i,\ell}B_{j,\ell}C_{k,\ell}$.
We write $\K(A,B,C)$ for the Kruskal product of the matrices $A,B,C$. 

\medskip
\noindent
{\em Trilinear form and bilinear map and defined by a tensor.}
Let $T\in\Ri^{m\times n\times p}$ be a tensor. 
For vectors $f\in\Ri^m$, $g\in\Ri^n$, and $h\in\Ri^p$, we define 
$T[f,g,h]=\sum_{i\in[m]}\sum_{j\in [n]}\sum_{k\in [p]}T_{i,j,k}f_ig_jh_k$
and say that $T[\cdot,\cdot,\cdot]$ is the {\em trilinear form} defined
by the tensor $T$. 
For vectors $u\in\Ri^n$, $v\in\Ri^p$, we define the vector $w=T[u,v]\in\Ri^m$
for all $i\in [m]$ by $w_i=\sum_{j\in [n]}\sum_{k\in [p]}T_{i,j,k}u_jv_k$
and say that $T[\cdot,\cdot]$ is the {\em bilinear map} defined
by the tensor $T$. 

\medskip
We collect select well-known properties of these product operations into 
the following lemma. 

\begin{Lem}[Product identities]
\label{lem:products}
The matrix, Kronecker, Hadamard, and Kruskal products satisfy:
\begin{enumerate}
\item[(i)]
For all 
$A\in\Ri^{m\times n}$, $B\in\Ri^{n\times p}$,
$C\in\Ri^{u\times v}$, and $D\in\Ri^{v\times w}$, we have
the matrix identity $(A\otimes C)(B\otimes D)=(AB)\otimes(CD)$.
\item[(ii)]
For all 
$A\in\Ri^{m\times r}$, $B\in\Ri^{n\times r}$, $C\in\Ri^{p\times r}$,
$A'\in\Ri^{m'\times r'}$, $B'\in\Ri^{n'\times r'}$, and 
$C'\in\Ri^{p'\times r'}$, we have
the tensor identity $\K(A,B,C)\otimes\K(A',B',C')=\K(A\otimes A',B\otimes B',C\otimes C')$.
\item[(iii)]
For all 
$A\in\Ri^{m\times r}$, $B\in\Ri^{n\times r}$, $C\in\Ri^{p\times r}$,
$u\in \Ri^n$, and $v\in \Ri^p$ with $T=\K(A,B,C)$, we have
the vector identity $T[u,v]=A((B^\top u)\odot(C^\top v))$.
\item[(iv)]
For all 
$A\in\Ri^{m\times r}$, $B\in\Ri^{n\times r}$, $C\in\Ri^{p\times r}$,
$f\in \Ri^m$, $g\in \Ri^n$, and $h\in \Ri^p$ with $T=\K(A,B,C)$, we have
the scalar identity $T[f,g,h]=(A^\top f)^\top((B^\top g)\odot(C^\top h))$.
\end{enumerate}
\end{Lem}
\begin{Proof}
Conclude from the definitions and commutativity of $\Ri$
that matching entries on the left-hand side and
the right-hand side of each identity (i)--(iv) are identical. 
\end{Proof}

\subsection{Tensor rank}
\label{sect:ranks}

Due to our choice to work in coordinates, 
it will be convenient to witness standard notions of tensor rank and
border rank as Kruskal products of matrices, in analogy with matrix 
rank witnessed as a matrix product.%
\footnote{Recall from elementary linear algebra that the rank of 
a nonzero matrix $C$ of shape $m\times n$ is the least positive 
integer $r$ such that there exist matrices $A$ and $B$ of shapes $m\times r$ 
and $r\times n$, respectively, with $C=AB$.}{}
We work with border rank using polynomials 
in one indeterminate---cf.~e.g.~Sch\"onhage~\cite{Schonhage1981}---which 
over the complex numbers is equivalent to the standard geometric definition
of border rank via secant varieties of a Segre variety, 
cf.~Landsberg~\cite{Landsberg2012} and 
Lehmkuhl and Lickteig~\cite{LehmkuhlL1989}.

\medskip
\noindent
{\em Tensor rank.}
The {\em rank} of a nonzero tensor $T\in\Ri^{m\times n\times p}$ is the least
positive integer $r$ such that there exist matrices $A\in\Ri^{m\times r}$,
$B\in\Ri^{n\times r}$, and $C\in\Ri^{p\times r}$ with $T=\K(A,B,C)$. 
A zero tensor has rank zero. 
We write $\Ra(T)$ for the rank of $T$.

\medskip
\noindent
{\em Border rank.}
Let us write $\Ri[\epsilon]$ for the ring of polynomials in 
the indeterminate $\epsilon$ with coefficients in $\Ri$. 
The {\em border rank} of a nonzero tensor $T\in\Ri^{m\times n\times p}$ is 
the least positive integer $r$ such that there exist matrices 
$A_\epsilon\in\Ri[\epsilon]^{m\times r}$,
$B_\epsilon\in\Ri[\epsilon]^{n\times r}$, 
$C_\epsilon\in\Ri[\epsilon]^{p\times r}$, 
a tensor $E_\epsilon\in\Ri[\epsilon]^{m\times n\times p}$, and a nonnegative integer
$d$ such that the {\em border decomposition} 
$\K(A_\epsilon,B_\epsilon,C_\epsilon)=\epsilon^d T+\epsilon^{d+1}E_\epsilon$ holds. We say that 
the $d$ is the {\em degree} and $E_\epsilon$ is the {\em error} tensor of 
the decomposition. A zero tensor has border rank zero. 
We write $\Rb(T)$ for the border rank of $T$.
It is immediate that $\Rb(T)\leq\Ra(T)$.

\subsection{Yates's algorithm}
\label{sect:yates}

Yates's algorithm~\cite{Yates1937} (cf.~also Knuth~\cite{Knuth1998}) 
enables us to multiply a vector with a Kronecker-product structured matrix 
fast by relying on sparse matrix-vector multiplication repeatedly. 
Suppose we are given as input the positive integer $p$, the matrices 
$A^{(\ell)}\in\Ri^{m_\ell\times n_\ell}$ for $\ell\in\{1,2,\ldots,p\}$, 
and a vector $x\in\Ri^{n_1n_2\cdots n_p}$. We are asked to output 
the vector $Ax\in \Ri^{m_1m_2\cdots m_p}$, where 
$A=A^{(1)}\otimes A^{(2)}\otimes\cdots\otimes A^{(p)}$. 
Writing $I_s$ for an $s\times s$ identity matrix, define the matrix
\[
A^{[\ell]}=I_{n_1}\otimes I_{n_2}\otimes\cdots\otimes I_{n_{\ell-1}}\otimes A^{(\ell)}\otimes I_{m_{\ell+1}}\otimes I_{m_{\ell+2}}\otimes \cdots\otimes I_{m_p}
\]
for $\ell\in\{1,2,\ldots,p\}$ and observe by Lemma~\ref{lem:products}(i) 
that $A=A^{[1]}A^{[2]}\cdots A^{[p]}$. Moreover, 
$A^{[\ell]}$ has at most 
$n_1n_2\cdots n_{\ell-1}n_{\ell}m_\ell m_{\ell+1}\cdots m_p$ nonzero
entries. Thus, since 
$Ax=A^{[1]}(A^{[2]}(\cdots (A^{[p]}x)\cdots))$, we can
compute $Ax$ from the given input in 
$\bO\bigl(\sum_{\ell=1}^p n_1n_2\cdots n_{\ell-1}n_{\ell}m_\ell m_{\ell+1}\cdots m_p\bigr)$ 
operations in~$\Ri$ by using sparse matrix-vector multiplication $p$ times.
For example, when $b=n_\ell$ and $r=m_\ell$ for all 
$\ell\in\{1,2,\ldots,p\}$, we obtain $\bO(\max(b,r)^{p+1} p)$ operations rather 
than the $\bO((br)^{p})$ operations which would result if we constructed
the matrix $A$ in explicit representation.

\subsection{Kronecker-product structured bilinear maps and trilinear forms}

Yates's algorithm enables the following fast algorithms for computing 
the bilinear map and the trilinear form defined by a tensor with 
Kronecker-product structure.
Let $T\in\Ri^{N\times N\times N}$ be a tensor with $T=S^{\otimes p}$ for
a {\em source} tensor $S\in\Ri^{b\times b\times b}$ and a positive integer $p$.
In particular, $N=b^p$. 

\medskip
\noindent
{\em Algorithms using a rank decomposition of the source tensor.}
Suppose that there exist three matrices $A,B,C\in\Ri^{b\times r}$ 
with $S=\K(A,B,C)$.
For two vectors $u,v\in\Ri^N$ and the matrices $A,B,C\in\Ri^{b\times r}$ given
as input, we observe that we have the following algorithm for computing
the vector $T[u,v]\in\Ri^N$ in total $\bO(\max(b,r)^{p+1}p)$ operations 
in $\Ri$ by relying on Yates's algorithm three times, interleaved
with one Hadamard product:
\begin{equation}
\label{eq:yates-bilinear}
\hat u\YL (B^\top)^{\otimes p}u\,,\quad
\hat v\YL (C^\top)^{\otimes p}v\,,\quad
\hat w\leftarrow \hat u\odot \hat v\,,\quad
     w\YL A^{\otimes p}\hat w\,;
\end{equation}
indeed, from Lemma~\ref{lem:products}(ii) we observe that
$T=S^{\otimes p}=\K(A^{\otimes p},B^{\otimes p},C^{\otimes p})$ 
and thus $w=T[u,v]$ by Lemma~\ref{lem:products}(iii). 
Similarly, for three vectors $f,g,h\in\Ri^N$ and the matrices 
$A,B,C\in\Ri^{b\times r}$ given as input, by Lemma~\ref{lem:products}(iv)
we observe that the following algorithm computes the trilinear form
$T[f,g,h]\in\Ri$ in total $\bO(\max(b,r)^{p+1}p)$ operations 
in $\Ri$:
\begin{equation}
\label{eq:yates-trilinear}
\hat f\YL (A^\top)^{\otimes p}f\,,\quad
\hat g\YL (B^\top)^{\otimes p}g\,,\quad
\hat h\YL (C^\top)^{\otimes p}h\,,\quad
T[f,g,h]\leftarrow \hat f^\top(\hat g\odot \hat h)\,.
\end{equation}

\medskip
\noindent
{\em Algorithms using a border decomposition of the source tensor.}
Suppose that there exist three matrices 
$A_\epsilon,B_\epsilon,C_\epsilon\in\Ri[\epsilon]^{b\times r}$,
a tensor $E_\epsilon\in\Ri[\epsilon]^{m\times n\times p}$, and 
a nonnegative integer $d$ with 
$\K(A_\epsilon,B_\epsilon,C_\epsilon)=\epsilon^d S+\epsilon^{d+1}E_\epsilon$.
Without loss of generality we may assume that no polynomial in the matrices
$A_\epsilon,B_\epsilon,C_\epsilon$ has degree more than $d$; indeed, otherwise
remove monomials of degree greater than $d$.
For a polynomial $q_\epsilon\in\Ri[\epsilon]$ and a nonnegative integer $k$, 
let us write $\{\epsilon^k\}q_\epsilon\in\Ri$ for the coefficient of the 
monomial $\epsilon^k$ in $q_\epsilon$, and extend the notation entrywise 
to vectors and matrices of such polynomials. 
The algorithms~\eqref{eq:yates-bilinear} 
and \eqref{eq:yates-trilinear} now extend to the algorithms
\begin{equation}
\label{eq:yates-bilinear-border}
\hat u_\epsilon\YL (B^\top_\epsilon)^{\otimes p}u\,,\quad
\hat v_\epsilon\YL (C^\top_\epsilon)^{\otimes p}v\,,\quad
\hat w_\epsilon\leftarrow \hat u_\epsilon\odot \hat v_\epsilon\,,\quad
     w_\epsilon\YL A^{\otimes p}_\epsilon\hat w_\epsilon\,,\quad
w \leftarrow \{\epsilon^{dp}\} w_\epsilon
\end{equation}
and
\begin{equation}
\label{eq:yates-trilinear-border}
\hat f_\epsilon\YL (A^\top_\epsilon)^{\otimes p}f\,,\quad 
\hat g_\epsilon\YL (B^\top_\epsilon)^{\otimes p}g\,,\quad
\hat h_\epsilon\YL (C^\top_\epsilon)^{\otimes p}h\,,\quad
T[f,g,h]\leftarrow \{\epsilon^{dp}\}\bigl(\hat f^\top_\epsilon(\hat g_\epsilon\odot \hat h_\epsilon)\bigr)\,,
\end{equation}
respectively. 
Correctness follows from $\K(A_\epsilon,B_\epsilon,C_\epsilon)=\epsilon^d S+\epsilon^{d+1}E_\epsilon$ and Lemma~\ref{lem:products}(ii,iii,iv) applied over
$\Ri[\epsilon]$ in place of $\Ri$. 
We immediately observe that all $\epsilon$-polynomials in the vectors
$\hat u_\epsilon,\hat v_\epsilon,\hat w_\epsilon,\hat f_\epsilon,\hat g_\epsilon,\hat g_\epsilon\in\Ri[\epsilon]^{r^p}$ 
have degree 
at most $\bO(dp)$, implying that the algorithms \eqref{eq:yates-bilinear-border}
and \eqref{eq:yates-trilinear-border} can be implemented so that they run 
in total $\bO(\max(b,r)^{p+1}d^2p^3)$ operations in~$\Ri$. 

\subsection{Fast subset convolution and three-way partitioning}
\label{sect:fast-subset-convolution}

This section derives fast subset convolution~\cite{BjorklundHKK2007} in the 
language of bilinear maps and trilinear forms defined by the Kronecker
power of an indicator tensor for three-way partitioning. 
Let~$\F$ be a field. In this section we assume 
that~$\Ri=\F$. We recall from \eqref{eq:three-way-partitioning} 
the {\em three-way partitioning} tensor 
\[
P=\left[\begin{array}{@{\,}c@{\ }|@{\ }c@{\,}}
\begin{array}{@{\,}c@{\,}c@{\,}}
0&1\\
1&0
\end{array} &
\begin{array}{@{\,}c@{\,}c@{\,}}
1&0\\
0&0
\end{array}
\end{array}\right]\,.
\]
It is well known that $\Rb(P)=2$; in particular, we have
$\K(A_\epsilon,B_\epsilon,C_\epsilon)=\epsilon P+\epsilon^2E_\epsilon$
with
\begin{equation}
\label{eq:three-way-partitioning-border-2}
A_\epsilon=\left[\begin{array}{@{\ }c@{\ \ }c@{\ }}
1&1\\
\epsilon&0\\
\end{array}\right]\,,\quad
B_\epsilon=\left[\begin{array}{@{\ }c@{\ \ }c@{\ }}
1&1\\
\epsilon&0\\
\end{array}\right]\,,\quad
C_\epsilon=\left[\begin{array}{@{\ }c@{\ \ }r@{\ }}
1&-1\\
\epsilon&0\\
\end{array}\right]\,,\quad
E_\epsilon=\left[\begin{array}{@{\,}c@{\ }|@{\ }c@{\,}}
\begin{array}{@{\,}c@{\ }c@{\,}}
0&0\\
0&1\\
\end{array} &
\begin{array}{@{\,}c@{\ }c@{\,}}
0&1\\
1&\epsilon\\
\end{array}
\end{array}\right]\,.
\end{equation}

Now let $n$ be a positive integer and associate each integer $x\in[2^n]$ 
through its binary representation $x=\sum_{j=0}^{n-1}2^jx_j$ 
for the bits $x_0,x_1,\ldots,x_{n-1}\in\{0,1\}$ with the unique
subset $X\subseteq [n]$ that satisfies $j\in X$ if and only if $x_j=1$ 
for all $j\in [n]$. With this association, for the tensor $T=P^{\otimes n}$ 
we observe that $T_{x,y,z}=1$ holds for $x,y,z\in [2^n]$ if and only if
the associated three subsets $X,Y,Z\in[n]$ satisfy $X\dcup Y\dcup Z=[n]$.

For two vectors $u,v\in\F^{2^n}$, define their {\em subset
convolution} $u*v\in\F^{2^n}$ for all $X\in 2^{[n]}$ by 
\[
(u*v)_X=\sum_{\substack{Y,Z\subseteq [n]\\Y\dcup Z=X}}u_Yv_Z
=\sum_{\substack{Y\subseteq X}}u_Yv_{X\setminus Y}\,.
\]
For $w\in\F^{2^n}$, define the {\em complement} $\bar w\in\F^{2^n}$ for
all $x\in[2^n]$ by $\bar w_x=w_{\bar x}$ where $\bar X=[n]\setminus X$.
We observe that $u*v=\overline{T[u,v]}$, and thus using 
\eqref{eq:three-way-partitioning-border-2} in \eqref{eq:yates-bilinear-border},
we can compute $u*v$ from given $u$ and $v$ in 
$\bO(2^nn^3)$ operations in the field $\F$. 

Let $\cF,\cG,\cH\subseteq 2^{[n]}$ be three set families. 
Define corresponding three indicator vectors $f,g,h\in\F^{2^n}$ by setting
$f_x=\iv{X\in\cF}$, $g_y=\iv{Y\in\cG}$, $h_z=\iv{Z\in\cH}$ for
all $x,y,z\in[2^n]$. 
We now have 
\[
T[f,g,h]=
\sum_{x,y,z\in [2^n]}T_{x,y,z}f_xg_yh_z=
|\{(X,Y,Z)\in\cF\times\cG\times\cH:X\dcup Y\dcup Z=[n]\}|_{\F}\,,
\]
where the subscript $\F$ indicates that the cardinality is reduced modulo 
the characteristic of $\F$. Using~\eqref{eq:three-way-partitioning-border-2}
in \eqref{eq:yates-trilinear-border}, we observe that we can compute
$T[f,g,h]$ from given $f,g,h$ in $\bO(2^nn^3)$ operations in the field $\F$. 
Working over the field
of rational numbers in particular, we observe that from $\cF,\cG,\cH$ given
as input, we can compute the number of triples 
$(X,Y,Z)\in\cF\times\cG\times\cH$ with $X\dcup Y\dcup Z=[n]$ in 
$\bO(2^nn^4)$ time.

\subsection{The asymptotic rank conjecture}

Let $\F$ be a field and let $S\in\F^{m\times n\times p}$ be a tensor.
Define the three {\em flattening} matrices $S^{[1]}\in \F^{m\times np}$,
$S^{[2]}\in \F^{n\times pm}$, and $S^{[3]}\in \F^{p\times mn}$ for
all $i\in [m]$, $j\in [n]$, and $k\in [p]$ by
$S^{[1]}_{i,jp+k}=S^{[2]}_{j,km+i}=S^{[3]}_{k,in+j}=S_{i,j,k}$.
Writing $\rk A$ for the rank of a matrix $A$, the tensor $S$
is {\em concise} if $\rk S^{[1]}=m$, $\rk S^{[2]}=n$, and $\rk S^{[3]}=p$.

Define the {\em support} 
$\supp S=\{(i,j,k)\in[m]\times [n]\times [p]:S_{i,j,k}\neq 0\}$. 
We say that the tensor $S$ is {\em tight} if there exist three
injective functions $\mu:[m]\rightarrow\Z$, 
$\nu:[n]\rightarrow\Z$, and 
$\pi:[p]\rightarrow\Z$ such that $\mu(i)+\nu(j)+\pi(k)=0$ for all 
$(i,j,k)\in \supp S$. 

We are now ready to restate Strassen's asymptotic rank 
conjecture~(Conjecture~\ref{conj:asymptotic-rank}; \cite[Conjecture~5.3]{Strassen1994}); we present the conjecture in a form 
studied by Conner {\em et al.}~\cite[Conjecture~1.4]{ConnerGLVW2021}
(for the field of complex numbers);
the lower bound $c^n\leq \Ra(S^{\otimes n})$ is immediate by conciseness
and the fact that matrix rank is multiplicative under Kroneckering.

\begin{Conj}[Asymptotic rank over $\F$; Strassen~\cite{Strassen1994}]
Let $S\in\F^{c\times c\times c}$ be concise and tight. 
Then, for all positive integers $n$ it holds that 
$c^n\leq \Ra(S^{\otimes n})\leq c^{n+o(n)}$.
\end{Conj}

While the asymptotic rank conjecture remains open, it is known that tensor 
rank is submultiplicative asymptotically. For example, over the complex numbers 
there exist tensors $S\in\F^{c\times c\times c}$ with $\Ra(S)\geq\Omega(c^2)$; 
 e.g.~\cite[Theorems~3.1.4.3 and 5.5.1.1]{Landsberg2012}. 
Yet, asymptotically the following theorem is implicit in Strassen's 
work~\cite[Proposition~3.6]{Strassen1988};
see~also~\cite[Proposition~2.12]{ChristandlVZ2021} and
\cite[Remark~2.1]{ConnerGLV2022}.
\begin{Thm}[Asymptotic submultiplicativity over $\F$; Strassen~\cite{Strassen1988}]
Let $S\in\F^{c\times c\times c}$. Then, for all positive integers $n$ it 
holds that $\Ra(S^{\otimes n})\leq c^{\frac{2\omega}{3}n+o(n)}$,
where $\omega$ is the exponent of matrix multiplication over $\F$.
\end{Thm}

\section{Opportunistic three-way partitioning}

\label{sect:three-way}

This section presents our main technical contribution.
We study the following {\em three-way partitioning} problem:
given three sets $\cF,\cG,\cH\subseteq 2^{[n]}$ as input, 
decide whether there exist sets $X\in\cF$, 
$Y\in\cG$, and $Z\in \cH$ such that $X\dcup Y\dcup Z=[n]$. 
For a nonnegative integer $k$, we say that the input $(\cF,\cG,\cH)$
is $k$-{\em bounded} if all the sets in $\cF,\cG,\cH$ have size at most $k$.

It is immediate that three-way partitioning can be solved in time $\bO^*(2^n)$
using fast subset convolution; cf.~Section~\ref{sect:fast-subset-convolution}.
Our objective in this section is the following lemma, which is 
Lemma~\ref{lem:main-three-way} restated here for convenience.

\begin{Lem}[Three-way partitioning under the asymptotic rank conjecture]
\label{lem:three-way}
Let\/ $\F$ be either a finite field or the field of complex numbers. 
If the asymptotic rank conjecture is true over\/~$\F$, then there
exist constants $\delta,\tau>0$ as well as a randomized algorithm that solves 
the three-way partitioning problem with high probability 
in $\bO((2-\delta)^n)$ time for
$(\frac{1}{3}+\tau)n$-bounded families of subsets.
\end{Lem}

This section presents a proof of Lemma~\ref{lem:three-way} 
when $\F$ is a finite field. The proof for the field of complex numbers is
postponed to Appendix~\ref{sect:complex}.

\subsection{Breaking a three-way partitioning tensor}

Let $\F$ be a finite field, and let $P$ be the three-way partitioning 
tensor \eqref{eq:three-way-partitioning} over $\F$. 
For concreteness and intuition, let us display the third Kronecker 
power of $P$, and disregard for the moment that we have underlined certain
$1$-entries in the tensor; we have:
\begin{equation}
\label{eq:p-powers}
\scriptsize
\begin{split}
P^{\otimes 3}&=\left[\begin{array}{@{\,}c@{\,}|@{\,}c@{\,}|@{\,}c@{\,}|@{\,}c@{\,}|@{\,}c@{\,}|@{\,}c@{\,}|@{\,}c@{\,}|@{\,}c@{\,}}
\begin{array}{@{\,}c@{\,}c@{\,}c@{\,}c@{\,}c@{\,}c@{\,}c@{\,}c@{\,}}
0&0&0&0&0&0&0&\underline{1}\\
0&0&0&0&0&0&1&0\\
0&0&0&0&0&1&0&0\\
0&0&0&0&1&0&0&0\\
0&0&0&1&0&0&0&0\\
0&0&1&0&0&0&0&0\\
0&1&0&0&0&0&0&0\\
\underline{1}&0&0&0&0&0&0&0
\end{array} &
\begin{array}{@{\,}c@{\,}c@{\,}c@{\,}c@{\,}c@{\,}c@{\,}c@{\,}c@{\,}}
0&0&0&0&0&0&1&0\\
0&0&0&0&0&0&0&0\\
0&0&0&0&1&0&0&0\\
0&0&0&0&0&0&0&0\\
0&0&1&0&0&0&0&0\\
0&0&0&0&0&0&0&0\\
1&0&0&0&0&0&0&0\\
0&0&0&0&0&0&0&0
\end{array} &
\begin{array}{@{\,}c@{\,}c@{\,}c@{\,}c@{\,}c@{\,}c@{\,}c@{\,}c@{\,}}
0&0&0&0&0&1&0&0\\
0&0&0&0&1&0&0&0\\
0&0&0&0&0&0&0&0\\
0&0&0&0&0&0&0&0\\
0&1&0&0&0&0&0&0\\
1&0&0&0&0&0&0&0\\
0&0&0&0&0&0&0&0\\
0&0&0&0&0&0&0&0
\end{array} &
\begin{array}{@{\,}c@{\,}c@{\,}c@{\,}c@{\,}c@{\,}c@{\,}c@{\,}c@{\,}}
0&0&0&0&1&0&0&0\\
0&0&0&0&0&0&0&0\\
0&0&0&0&0&0&0&0\\
0&0&0&0&0&0&0&0\\
1&0&0&0&0&0&0&0\\
0&0&0&0&0&0&0&0\\
0&0&0&0&0&0&0&0\\
0&0&0&0&0&0&0&0
\end{array} &
\begin{array}{@{\,}c@{\,}c@{\,}c@{\,}c@{\,}c@{\,}c@{\,}c@{\,}c@{\,}}
0&0&0&1&0&0&0&0\\
0&0&1&0&0&0&0&0\\
0&1&0&0&0&0&0&0\\
1&0&0&0&0&0&0&0\\
0&0&0&0&0&0&0&0\\
0&0&0&0&0&0&0&0\\
0&0&0&0&0&0&0&0\\
0&0&0&0&0&0&0&0
\end{array} &
\begin{array}{@{\,}c@{\,}c@{\,}c@{\,}c@{\,}c@{\,}c@{\,}c@{\,}c@{\,}}
0&0&1&0&0&0&0&0\\
0&0&0&0&0&0&0&0\\
1&0&0&0&0&0&0&0\\
0&0&0&0&0&0&0&0\\
0&0&0&0&0&0&0&0\\
0&0&0&0&0&0&0&0\\
0&0&0&0&0&0&0&0\\
0&0&0&0&0&0&0&0
\end{array} &
\begin{array}{@{\,}c@{\,}c@{\,}c@{\,}c@{\,}c@{\,}c@{\,}c@{\,}c@{\,}}
0&1&0&0&0&0&0&0\\
1&0&0&0&0&0&0&0\\
0&0&0&0&0&0&0&0\\
0&0&0&0&0&0&0&0\\
0&0&0&0&0&0&0&0\\
0&0&0&0&0&0&0&0\\
0&0&0&0&0&0&0&0\\
0&0&0&0&0&0&0&0
\end{array} &
\begin{array}{@{\,}c@{\,}c@{\,}c@{\,}c@{\,}c@{\,}c@{\,}c@{\,}c@{\,}}
\underline{1}&0&0&0&0&0&0&0\\
0&0&0&0&0&0&0&0\\
0&0&0&0&0&0&0&0\\
0&0&0&0&0&0&0&0\\
0&0&0&0&0&0&0&0\\
0&0&0&0&0&0&0&0\\
0&0&0&0&0&0&0&0\\
0&0&0&0&0&0&0&0
\end{array} 
\end{array}\right]\,.
\end{split}
\end{equation}
An immediate induction yields that the tensor $P^{\otimes n}$ has shape 
$2^n\times 2^n\times 2^n$ and a support of size~$3^n$. Furthermore, it
is immediate that $P^{\otimes n}$ has $\Ra(P^{\otimes n})\geq 2^n$ because 
the $2^n\times 2^n$ identity matrix occurs as a matrix slice. Thus, 
the tensor $P^{\otimes n}$ in itself will not yield sub-$2^n$-time algorithms, 
even under the asymptotic rank conjecture. 

We will follow a strategy of ``breaking and randomizing''---%
cf.~Karppa and Kaski~\cite{KarppaK2019}, who introduced the technique
for matrix multiplication tensors---%
the tensor $P^{\otimes n}$. Namely, we will ``break'' the tensor by 
setting some of its $1$-entries to $0$-entries, thereby enabling---under the 
asymptotic rank conjecture in our case---lower tensor rank and thus
a sub-$2^n$-time broken algorithm with false negatives caused by the
introduced $0$-entries. We will then use randomization with independent 
random permutations of the universe $[n]$ to obtain an algorithm that 
is correct with high probability, 
still retaining an overall sub-$2^n$ running time in the process.
In contrast to a matrix multiplication tensor, which admits a group of
permutation automorphisms that acts transitively on the $1$-entries 
of the tensor~\cite{KarppaK2019}, the tensor $P^{\otimes n}$ does not
have this property, which requires further care in breaking the tensor 
and in analysing the randomization for restoring correctness. 

Let us work with the $8\times 8\times 8$ third power $P^{\otimes 3}$ 
illustrated above in~\eqref{eq:p-powers}. First, 
transform the $1$-entries at 
positions $(0,0,7)$, $(0,7,0)$, and $(7,0,0)$ of $P^{\otimes 3}$
into $0$-entries; these $1$-entries are underlined in \eqref{eq:p-powers}
for illustration. Next, cut out from the tensor that results 
the three $8\times 8$ all-zero-matrix slices---that is, cut out all entries 
with at least one coordinate equal to $7$ from the tensor---to obtain 
the $7\times 7\times 7$ tensor
\begin{equation}
\label{eq:q-tensor}
\small
Q=\left[\begin{array}{@{\,}c@{\,}|@{\,}c@{\,}|@{\,}c@{\,}|@{\,}c@{\,}|@{\,}c@{\,}|@{\,}c@{\,}|@{\,}c@{\,}}
\begin{array}{@{\,}c@{\,}c@{\,}c@{\,}c@{\,}c@{\,}c@{\,}c@{\,}}
0&0&0&0&0&0&0\\
0&0&0&0&0&0&1\\
0&0&0&0&0&1&0\\
0&0&0&0&1&0&0\\
0&0&0&1&0&0&0\\
0&0&1&0&0&0&0\\
0&\underline{1}&\underline{0}&\underline{0}&\underline{0}&\underline{0}&\underline{0}
\end{array} &
\begin{array}{@{\,}c@{\,}c@{\,}c@{\,}c@{\,}c@{\,}c@{\,}c@{\,}}
0&0&0&0&0&0&1\\
0&0&0&0&0&0&0\\
0&0&0&0&1&0&0\\
0&0&0&0&0&0&0\\
0&0&1&0&0&0&0\\
0&0&0&0&0&0&0\\
1&0&0&0&0&0&\underline{0}
\end{array} &
\begin{array}{@{\,}c@{\,}c@{\,}c@{\,}c@{\,}c@{\,}c@{\,}c@{\,}}
0&0&0&0&0&1&0\\
0&0&0&0&1&0&0\\
0&0&0&0&0&0&0\\
0&0&0&0&0&0&0\\
0&1&0&0&0&0&0\\
1&0&0&0&0&0&0\\
0&0&0&0&0&0&0
\end{array} &
\begin{array}{@{\,}c@{\,}c@{\,}c@{\,}c@{\,}c@{\,}c@{\,}c@{\,}}
0&0&0&0&1&0&0\\
0&0&0&0&0&0&0\\
0&0&0&0&0&0&0\\
0&0&0&0&0&0&0\\
1&0&0&0&0&0&0\\
0&0&0&0&0&0&0\\
0&0&0&0&0&0&0
\end{array} &
\begin{array}{@{\,}c@{\,}c@{\,}c@{\,}c@{\,}c@{\,}c@{\,}c@{\,}}
0&0&0&1&0&0&0\\
0&0&1&0&0&0&0\\
0&1&0&0&0&0&0\\
1&0&0&0&0&0&0\\
0&0&0&0&0&0&0\\
0&0&0&0&0&0&0\\
0&0&0&0&0&0&0
\end{array} &
\begin{array}{@{\,}c@{\,}c@{\,}c@{\,}c@{\,}c@{\,}c@{\,}c@{\,}}
0&0&1&0&0&0&0\\
0&0&0&0&0&0&0\\
1&0&0&0&0&0&0\\
0&0&0&0&0&0&0\\
0&0&0&0&0&0&0\\
0&0&0&0&0&0&0\\
0&0&0&0&0&0&0
\end{array} &
\begin{array}{@{\,}c@{\,}c@{\,}c@{\,}c@{\,}c@{\,}c@{\,}c@{\,}}
0&1&0&0&0&0&0\\
1&0&0&0&0&0&0\\
0&0&0&0&0&0&0\\
0&0&0&0&0&0&0\\
0&0&0&0&0&0&0\\
0&0&0&0&0&0&0\\
0&0&0&0&0&0&0
\end{array} 
\end{array}\right]\,.
\end{equation}

\begin{Lem}[Properties of the tensor $Q$]
\label{lem:q-properties}
The tensor $Q$ in \eqref{eq:q-tensor} is concise and tight as well as has
support size $24$. In particular, $Q$ meets the conditions of the asymptotic
rank conjecture.
\end{Lem}
\begin{Proof}
Support size and conciseness (over any field $\F$) are immediate as witnessed 
by the underlined columns in~\eqref{eq:q-tensor} as well as
$Q^{[1]}=Q^{[2]}=Q^{[3]}$.
To establish tightness, we observe
that $P$ is tight. Indeed, we have $\supp P=\{(0,0,1),(0,1,0),(1,0,0)\}$
and we can take $\mu(0)=-1$, $\mu(1)=1$, $\nu(0)=-1$, $\nu(1)=1$, 
$\pi(0)=0$, $\pi(1)=2$. Strassen~\cite[\S5]{Strassen1991} shows that 
tightness persists under taking of Kronecker products, implying that
$P^{\otimes 3}$ is tight. Since $Q$ is obtained from $P^{\otimes 3}$ by
restricting its support, $Q$ is tight.
\end{Proof}

Let $\rho>0$ be a constant whose precise value we will fix later. 
Since $Q$ is concise and tight over the finite field
$\F$ by Lemma~\ref{lem:q-properties}, 
under the asymptotic rank conjecture
there exists positive integer constant $d$ such that
$7^d\leq \Ra(Q^{\otimes d})\leq 7^{(1+\rho)d}$. In particular,
by definition of tensor rank, there exists a positive integer 
$7^d\leq r\leq 7^{(1+\rho)d}$ and three matrices 
$A_\Qm,B_\Qm,C_\Qm\in\F^{7^d\times r}$ with
\begin{equation}
\label{eq:q-rank}
\K(A_\Qm,B_\Qm,C_\Qm)=Q^{\otimes d}
\end{equation}
over the finite field $\F$. 
Let $\epsilon$ be a polynomial indeterminate and 
let $A_\epsilon,B_\epsilon,C_\epsilon\in\F[\epsilon]^{2\times 2}$
be the three matrices in~\eqref{eq:three-way-partitioning-border-2}.

\subsection{The broken algorithm and its analysis}
\label{sect:broken-algorithm}

Let $n$ be a nonnegative integer. We may assume that $n=p+3dq$ for two 
nonnegative integers $p$ and $q$. Let $\cF,\cG,\cH\subseteq 2^{[n]}$ be given as
input. We say that a subset $X\subseteq [n]$ is {\em bad} if there
exists a $j\in [dq]$ with $\{3j,3j+1,3j+2\}\subseteq X$; otherwise $X$ 
is {\em good}. 
A three-way partition $X\dcup Y\dcup Z=[n]$ with 
$(X,Y,Z)\in\cF\times\cG\times\cH$ is {\em good} if each of the sets 
$X,Y,Z$ is good; otherwise the partition is {\em bad}. 
We present a randomized algorithm that decides whether 
$\cF\times\cG\times\cH$ contains a good three-way partition with
high probability. 

Towards this end, recalling that we identify each subset $X\subseteq [n]$
with the unique $n$-bit binary integer $x=\sum_{j\in [n]}2^j x_j$ with
$x_j=\iv{j\in X}$ for all $j\in [n]$, it will be convenient to 
introduce a mixed-base integer representation restricted to the 
good subsets only. First, observe that $X\subseteq[n]$ is good if and only if 
there does {\em not} exist a $j\in [dq]$ with $x_{3j}=x_{3j+1}=x_{3j+2}=1$. 
Accordingly, when $X\subseteq[n]$ is good, we have 
$0\leq x_{3j}+2x_{3j+1}+4x_{3j+2}\leq 6$ for all $j\in [q]$. 
In particular, it is immediate that there are exactly $2^p7^{dq}$ good
subsets in $2^{[n]}$, and we can set up a bijection between the set of all 
good subsets and the set $[2^p7^{dq}]$ as follows. Associate with a good subset 
$X\subseteq[n]$ the unique integer $\underline{x}\in [2^p7^{dq}]$ defined by
\begin{equation}
\label{eq:x-underline}
\underline{x}\ =\ 7^{dq}\!\!\!\!\!\sum_{j\in [n]\setminus[3dq]}\!\!\!\!\!2^{j-3dq}x_j\ +\ \sum_{j\in[dq]}7^j(x_{3j}+2x_{3j+1}+4x_{3j+2})\,.
\end{equation}

We are now ready to describe the algorithm. First, the algorithm removes all
the bad sets in each of the families $\cF,\cG,\cH$ given as input; let us
write $\underline{\cF},\underline{\cG},\underline{\cH}$ for the set families
so obtained. Second, the algorithm constructs
an extension field $\E\supseteq\F$ of degree $n$ of the finite field $\F$ 
by finding an $\F$-irreducible polynomial of degree $n$; 
cf.~\cite[\S14]{vonzurGathenG2013}.
Third, for all 
$\underline{x},\underline{y},\underline{z}\in [2^p7^{dq}]$, 
the algorithm draws three independent uniform random elements 
$\phi_{\underline{x}},\psi_{\underline{y}},\chi_{\underline{z}}\in\E$.
Fourth, the algorithm constructs the vectors 
$\underline{f},\underline{g},\underline{h}\in\E^{2^p7^{dq}}$ defined
for all $\underline{x},\underline{y},\underline{z}\in [2^p7^{dq}]$ by
\begin{equation}
\label{eq:fgh-underline}
\underline{f}_{\underline{x}}=\iv{X\in\underline{\cF}}\phi_{\underline{x}}\,,\quad
\underline{g}_{\underline{y}}=\iv{Y\in\underline{\cG}}\psi_{\underline{y}}\,,\quad
\underline{h}_{\underline{z}}=\iv{Z\in\underline{\cH}}\chi_{\underline{z}}\,.
\end{equation}
Fifth, using Yates's algorithm (Sect.~\ref{sect:yates}) over the 
polynomial ring 
$\Ri=\E[\epsilon]\supseteq\E\supseteq\F$ in each case, the algorithm 
computes the three matrix-vector products
\begin{equation}
\label{eq:broken-yates}
\begin{split}
\tilde f_\epsilon&\YL ((A_\epsilon^\top)^{\otimes p}\otimes (A_\Qm^\top)^{\otimes q})\underline f\,,\\
\tilde g_\epsilon&\YL ((B_\epsilon^\top)^{\otimes p}\otimes (B_\Qm^\top)^{\otimes q})\underline g\,,\\
\tilde h_\epsilon&\YL ((C_\epsilon^\top)^{\otimes p}\otimes (C_\Qm^\top)^{\otimes q})\underline h\,.
\end{split}
\end{equation}
Finally, the algorithm outputs the $\E$-element
\begin{equation}
\label{eq:broken-output}
\Gamma\leftarrow\{\epsilon^p\}(\tilde f_\epsilon^\top(\tilde g_\epsilon\odot\tilde  h_\epsilon))\,.
\end{equation}

We now proceed to analyze the broken algorithm for its 
limited correctness and running time.

\begin{Lem}[The broken algorithm is correct with high probability for good
three-way partitions]
\label{lem:broken-on-good}
The output of the broken algorithm satisfies\/
$\Gamma=T[\underline{f},\underline{g},\underline{h}]$ for the
tensor $T=P^{\otimes p}\otimes Q^{\otimes dq}$. 
Moreover, 
\begin{enumerate}
\item[(i)]
if the input
has a good three-way partition, then $\Gamma\neq 0$ with probability at least
$1-\bO\bigl(|\F|^{-n}\bigr)$; 
\item[(ii)]
if the input
does not have
a good three-way partition, then $\Gamma=0$ with probability $1$.
\end{enumerate}
\end{Lem}
\begin{Proof}
To analyse the broken algorithm, let us view 
$\phi_{\underline{x}},\psi_{\underline{y}},\chi_{\underline{z}}$
for $\underline{x},\underline{y},\underline{z}\in[2^p7^{dq}]$ 
not as independent uniform random elements of $\E$ (as drawn in the third 
step of the algorithm) but rather as formal polynomial indeterminates, 
to which we will eventually assign the chosen random values. 
Accordingly, let us first work over the multivariate polynomial ring
\begin{equation}
\label{eq:ring-analysis}
\Ri=\E[\epsilon][\phi_{\underline{x}},\psi_{\underline{y}},\chi_{\underline{z}}:\underline{x},\underline{y},\underline{z}\in[2^p7^{dq}]]\,.
\end{equation}
From Lemma~\ref{lem:products}(iv), \eqref{eq:broken-output}, and 
\eqref{eq:broken-yates} we have
\begin{equation}
\label{eq:t-epsilon-form}
T_\epsilon[\underline{f},\underline{g},\underline{h}]
=\tilde f_\epsilon^\top(\tilde g_\epsilon\odot\tilde  h_\epsilon)
\end{equation}
for the tensor $T_\epsilon=\K(A,B,C)$ with
\begin{equation}
\label{eq:t-epsilon-abc}
A=A_\epsilon^{\otimes p}\otimes A_\Qm^{\otimes q}\,,\qquad
B=B_\epsilon^{\otimes p}\otimes B_\Qm^{\otimes q}\,,\qquad
C=C_\epsilon^{\otimes p}\otimes C_\Qm^{\otimes q}\,.
\end{equation}
From \eqref{eq:t-epsilon-abc}, Lemma~\ref{lem:products}(ii),
\eqref{eq:three-way-partitioning-border-2}, and \eqref{eq:q-rank} 
we thus have
\begin{equation}
\label{eq:t-epsilon}
T_\epsilon
=(\epsilon P+\epsilon^2E_\epsilon)^{\otimes p}\otimes Q^{\otimes q}
=\epsilon^p P^{\otimes p}\otimes Q^{\otimes q}+\epsilon^{p+1}\bar E_\epsilon
\end{equation}
for a tensor 
$\bar E_\epsilon\in\Ri^{2^p7^{dq}\times 2^p7^{dq}\times 2^p7^{dq}}$.
From \eqref{eq:t-epsilon}, \eqref{eq:t-epsilon-form}, 
and \eqref{eq:broken-yates} it thus follows that
\begin{equation}
\label{eq:gamma-t}
\Gamma
=\{\epsilon^p\}T_\epsilon[\underline{f},\underline{g},\underline{h}]
=T[\underline{f},\underline{g},\underline{h}]
\end{equation}
for the tensor $T=P^{\otimes p}\otimes Q^{\otimes q}$.
It follows immediately from \eqref{eq:x-underline}, 
\eqref{eq:three-way-partitioning}, \eqref{eq:q-tensor}, and the definition
of the Kronecker product for tensors that this tensor $T$ satisfies 
for all good subsets $X,Y,Z\subseteq[n]$ the identity
\begin{equation}
\label{eq:t-good-partitioning}
T_{\underline{x},\underline{y},\underline{z}}=\iv{X\dcup Y\dcup Z=[n]}\,.
\end{equation}
From \eqref{eq:gamma-t}, \eqref{eq:t-good-partitioning}, 
and \eqref{eq:fgh-underline} we thus have
\begin{equation}
\label{eq:gamma-polynomial}
\Gamma
=
\sum_{\underline{x},\underline{y},\underline{z}\in[2^p7^{dq}]}
T_{\underline{x},\underline{y},\underline{z}}
\underline{f}_{\underline{x}}
\underline{g}_{\underline{y}}
\underline{h}_{\underline{z}}
=
\sum_{\substack{%
(X,Y,Z)\in\underline{\cF}\times \underline{\cG}\times \underline{\cH}\\
X\dcup Y\dcup Z=[n]}}
\phi_{\underline{x}}
\psi_{\underline{y}}
\chi_{\underline{z}}\,.
\end{equation}
That is, recalling \eqref{eq:ring-analysis} and that we view 
$\phi_{\underline{x}},\psi_{\underline{y}},\chi_{\underline{z}}$
for $\underline{x},\underline{y},\underline{z}\in[2^p7^{dq}]$
as polynomial indeterminates for the ring $\Ri$, we observe 
from \eqref{eq:gamma-polynomial} that $\Gamma$ is the identically zero
polynomial in these indeterminates if and only if there exists 
no three-way partition in 
$\underline{\cF}\times \underline{\cG}\times \underline{\cH}$; otherwise
$\Gamma$ is a polynomial of degree $3$. 
Now, as is done in the broken algorithm, substituting the independent 
uniform random elements of $\E$ in place of the indeterminates
$\phi_{\underline{x}},\psi_{\underline{y}},\chi_{\underline{z}}$
for $\underline{x},\underline{y},\underline{z}\in[2^p7^{dq}]$,
we observe that $\Gamma=0$ over $\E$ with probability $1$ if there exists 
no three-way partition in 
$\underline{\cF}\times \underline{\cG}\times \underline{\cH}$; otherwise,
$\Gamma\neq 0$ over $\E$ 
with probability at least $1-3/|\E|=1-\bO\bigl(|\F|^{-n}\bigr)$
by the DeMillo-Lipton-Schwartz-Zippel 
Lemma~\cite{DeMilloL1978,Schwartz1980,Zippel1979}.
\end{Proof}

Let us now analyze the running time of the broken algorithm. 
Since $|\F|\geq 2$ is a constant, we have that the arithmetic operations
(addition, negation, multiplication, inversion) in the extension field 
$\E\supseteq\F$ run in time bounded by a polynomial in $n$. The construction
of the extension field $\E$ itself can be assumed to be a one-time 
(randomized) operation that runs in time polynomial in $n$ and fails with
probability at most $\bO(|\F|^{-n})$; cf.~\cite[\S14]{vonzurGathenG2013}.
Thus, the first four steps of the algorithm run 
in time $\bO\bigl((|\cF|+|\cG|+|\cH|+2^p7^{dq})\poly(n)\bigr)$.
In particular, for $0<\tau<1/6$
we have $|\cF|,|\cG|,|\cH|\leq 2^{H(1/3+\tau)n}$, where 
$H(\lambda)=-\lambda\log_2\lambda-(1-\lambda)\log_2(1-\lambda)$ is the 
binary entropy function; cf.~\cite[\S1]{Jukna2011}.
Since every $\epsilon$-polynomial in $\E[\epsilon]$ that the algorithm
works with has degree at most $3p$ with $p\leq n=p+3dq$, and recalling that the
matrices $A_\epsilon,B_\epsilon,C_\epsilon$ and $A_\Qm,B_\Qm,C_\Qm$
have shapes $2\times 2$ and $7^d\times r$ with $7^d\leq r\leq 7^{(1+\rho)d}$,
respectively, from the analysis of Yates's algorithm (Sect.~\ref{sect:yates}) 
it follows that the fifth step as well as the final step of the algorithm run 
in $\bO\bigl(2^p 7^{(1+\rho)dq}\poly(n)\bigr)$ time. 
Thus, the broken algorithm runs in total 
$\bO\bigl((2^{H(1/3+\tau)n}+2^p 7^{(1+\rho)dq})\poly(n)\bigr)$ time
for constants $\rho>0$ and $0<\tau<1/6$ as well as nonnegative integers 
$p$ and $q$ with $p+3dq=n$.

\subsection{Randomizing the input to the broken algorithm}
\label{sect:broken-randomize}

We will now correct the broken algorithm so that it correctly 
decides with high probability whether a given input 
$\cF,\cG,\cH\subseteq 2^{[n]}$
admits a three-way partition $X\dcup Y\dcup Z=[n]$ with $X\in\cF$, 
$Y\in\cG$, $Z\in\cH$. Our approach is simply to repeat the
following $s$ times, with a precise value of $s$ to be fixed later. 
First, draw a uniform random permutation $\pi:[n]\rightarrow[n]$.
For a set $S\subseteq[n]$, let us write $\pi(S)=\{\pi(i):i\in S\}$.
Second, construct the permuted input 
\[
\pi(\cF)=\{\pi(X):X\in\cF\}\,,\quad
\pi(\cG)=\{\pi(Y):Y\in\cG\}\,,\quad
\pi(\cH)=\{\pi(Z):Z\in\cH\}\,.\quad
\]
Third, run the broken algorithm with the permuted input 
$\pi(\cF),\pi(\cG),\pi(\cH)$. If the broken algorithm outputs a nonzero 
value, assert that a three-way partition has been found and stop;
otherwise continue with the next repeat, or assert that no three-way
partition exists and stop if all the $s$ repeats have been done. 

Recall from the previous section that we have $n=p+3dq$ for nonnegative 
integers $p$ and $q$ and that a set $X\subseteq [n]$ is {\em bad} if
there exists a $j\in[dq]$ with $\{3j,3j+1,3j+2\}\subseteq X$; otherwise
$X$ is {\em good}. 

\begin{Lem}[Probability of obtaining a good partition under random permutation]
\label{lem:good-probability}
Let $X,Y,Z\subseteq[n]$ with $|X|,|Y|,|Z|\leq (\frac{1}{3}+\tau)n$. 
Suppose that $\pi:[n]\rightarrow[n]$ is a uniform random permutation and 
that $dq\leq\sigma n$ for some constant 
$0<\sigma<\frac{1}{12}$. Then, 
\begin{equation}
\label{eq:good-probability}
\Pr_\pi\bigl(\text{each of $\pi(X)$, $\pi(Y)$, $\pi(Z)$ is good\/}\bigr)
\geq \biggl(1-\frac{3(\frac{1}{3}+\tau)^3}{(1-3\sigma)^3}\biggr)^{\sigma n}\,.
\end{equation}
\end{Lem}
\begin{Proof}
Let us write $\pi^{-1}$ for the inverse permutation of $\pi$. 
Let $j\in [dq]$ and let $f:[3j]\rightarrow [n]$ be injective. 
Let us write $\bad_{X,j}(\pi)$ for the event 
$\{\pi^{-1}(3j),\pi^{-1}(3j+1),\pi^{-1}(3j+2)\}\subseteq X$.
Let us write $\pi^{-1}|_{[3j]}$ for the restriction of $\pi^{-1}$ 
to the domain $[3j]$. We have the 
following uniform upper bound on the conditional probability
\[
\Pr_\pi\bigl(\bad_{X,j}(\pi)\big|\pi^{-1}|_{[3j]}=f\bigr)
=\frac{\binom{|X\setminus f([3j])|}{3}}{\binom{n-3j}{3}}
\leq \frac{|X|^3}{(n-3dq)^3}
\leq \frac{(\frac{1}{3}+\tau)^3}{(1-3\sigma)^3}\,.
\]
Thus, by the union bound we have
\[
\Pr_\pi\bigl(\bad_{X,j}(\pi)\cup\bad_{Y,j}(\pi)\cup\bad_{Z,j}(\pi)\big|\pi^{-1}|_{[3j]}=f\bigr)\leq \frac{3(\frac{1}{3}+\tau)^3}{(1-3\sigma)^3}\,.
\]
Writing $\good_j(\pi)$ for the complement of the event 
$\bad_{X,j}(\pi)\cup\bad_{Y,j}(\pi)\cup\bad_{Z,j}(\pi)$, we consequently have
\[
\Pr_\pi\bigl(\good_j(\pi)\big|\pi^{-1}|_{[3j]}=f\bigr)\geq 
1-\frac{3(\frac{1}{3}+\tau)^3}{(1-3\sigma)^3}\,.
\]
Now let us recall that for an event $E$ and pairwise disjoint events 
$F_1,F_2,\ldots,F_m$ it holds that $\Pr(E|F_i)\geq\beta$ for 
all $1\leq i\leq m$ implies $\Pr(E|\cup_{i=1}^mF_i)\geq\beta$. 
Since the event $\cap_{\ell\in [j]}\good_\ell(\pi)$ is a 
disjoint union of some nonempty set of the events $\pi^{-1}|_{[3j]}=f$, 
we have
\begin{equation}
\label{eq:prlb}
\Pr_\pi\bigl(\good_j(\pi)\big|\cap_{\ell\in [j]}\good_\ell(\pi)\bigr)\geq 
1-\frac{3(\frac{1}{3}+\tau)^3}{(1-3\sigma)^3}\,.
\end{equation}
Recalling that $0<\tau<1/6$, by our choice of $\sigma<\frac{1}{12}<(1-(3/8)^{1/3})/3$ we have that
the right-hand side of \eqref{eq:prlb} is positive. 
The lower bound \eqref{eq:good-probability} thus follows by chaining
the conditional probabilities \eqref{eq:prlb} over $j\in [dq]$.
\end{Proof}

We are now ready to analyse the success probability of the randomized
broken algorithm. Indeed, recall that we run $s$ independent repeats of 
the broken algorithm on input $\pi(\cF),\pi(\cG),\pi(\cH)$, where
each repeat uses an independent random $\pi$ and thus yields a good
three-way partition from any existing three-way partition in the 
master input $\cF,\cG,\cH$ with probability controlled 
by~\eqref{eq:good-probability}. 
Let us abbreviate 
$\theta=\bigl(1-\frac{3(\frac{1}{3}+\tau)^3}{(1-3\sigma)^3}\bigr)^{\sigma}$
for convenience, recalling from \eqref{eq:good-probability}
that the probability that $\pi$ succeeds in producing 
a good three-way partition $\pi(X)\dcup\pi(Y)\dcup\pi(Z)=[n]$
from an arbitrary fixed three-way partition $X\dcup Y\dcup Z=[n]$
in the input $\cF,\cG,\cH$ is then at least $\theta^n$. 

With foresight, let us run exactly 
$s=\lceil\theta^{-n}n\rceil$ independent repeats. 
Then, recalling that $1+\xi\leq \exp(\xi)$ for all real $\xi$, 
we have 
\[
\Pr(\text{at least one of the $s$ repeats succeeds})\geq 
1-(1-\theta^n)^s\geq
1-\exp(-\theta^ns)\geq
1-\exp(-n)\,.
\]
Assuming that a three-way partition exists in the input $\cF,\cG,\cH$, and 
conditioning that at least one of the $s$ repeats succeeds to yield
a good three-way partition to $\pi(\cF),\pi(\cG),\pi(\cH)$, 
from Lemma~\ref{lem:broken-on-good} we know that the broken algorithm discovers
the good three-way partition on such a repeat with probability at least
$1-\exp(-\Omega(n))$. We conclude that the overall randomized algorithm 
(i) asserts a three-way partition
with probability at least $1-\exp(-\Omega(n))$ if a three-way partition exists 
in the input; and
(ii) asserts that no three-way partition exists in the input 
with probability $1$ if no three-way partition exists in the input.

It remains to parameterize and analyse the running time of 
the overall algorithm. 
Since we run $s=\lceil\theta^{-n}n\rceil$ independent repeats of the 
broken algorithm with 
$\theta=\bigl(1-\frac{3(\frac{1}{3}+\tau)^3}{(1-3\sigma)^3}\bigr)^{\sigma}$
and $dq\leq\sigma n$ for a constant $0<\sigma<\frac{1}{12}$, 
and each repeat runs in 
$\bO\bigl((2^{H(1/3+\tau)n}+2^p 7^{(1+\rho)dq})\poly(n)\bigr)$ time
for constants $\rho>0$ and $0<\tau<1/6$ as well as nonnegative integers 
$p$ and $q$ with $p+3dq=n$, we obtain the total running time
\[
\bO\biggl(
\biggl(2^{H(1/3+\tau)n}+
2^p 7^{(1+\rho)dq}\biggr)\biggl(1-\frac{3(\frac{1}{3}+\tau)^3}{(1-3\sigma)^3}\biggr)^{-\sigma n}\poly(n)
\biggr)\,.
\]
Taking $q=\lfloor \sigma n/d\rfloor$ and $p=n-3dq$, we thus obtain
\[
\bO\biggl(
\biggl(
\biggl(2^{H(1/3+\tau)}\biggl(1-\frac{3(\frac{1}{3}+\tau)^3}{(1-3\sigma)^3}\biggr)^{-\sigma}\biggr)^n+
\biggl(
2^{(1-3\sigma)} 7^{(1+\rho)\sigma}
\biggl(1-\frac{3(\frac{1}{3}+\tau)^3}{(1-3\sigma)^3}\biggr)^{-\sigma}\biggr)^n\biggr)\poly(n)
\biggr)\,.
\]
Taking $\sigma=\tau=\rho=\frac{1}{1000}$, we conclude that the running
time is $\bO((2-\delta)^n)$ for $\delta=\frac{1}{100000}$. 
This completes the proof of Lemma~\ref{lem:three-way} when $\F$ is 
a finite field. The proof when $\F$ is the field of complex numbers
appears in Appendix~\ref{sect:complex}.

\section{Covering and partitioning}
\label{sect:covering-and-partitioning}

This section proves Theorem~\ref{thm:main} using 
Lemma~\ref{lem:main-three-way}. 
Namely, we present a combinatorial reduction from the 
$k$-\textsc{Set Cover} problem (Problem~\ref{prob:set-cover}) 
first to the 
$k$-\textsc{Set Partitioning} problem---that is, the $k$-\textsc{Set Cover}
problem with the further requirement that the sets 
$S_{j_1},S_{j_2},\ldots,S_{j_{t'}}$ 
in a solution should be pairwise disjoint---and then to the three-way 
partitioning problem. 

We begin by recalling an easy relationship between $k$-\textsc{Set Cover}
and $k$-\textsc{Set Partitioning}.
For a set family $\cF$, let us write $\down\cF$ 
the {\em down-closure} of $\cF$ consisting of all sets $X$ such that
there exists an $Y\in \cF$ with $X\subseteq Y$.
We say that $\cF$ is {\em down-closed} if $\cF=\down\cF$.

\begin{Lem}[Reducing $k$-\textsc{Set Cover} to $k$-\textsc{Set Partitioning}]
\label{lem:covering-to-partitioning}
Suppose that there exists a constant $\beta>0$ and an algorithm that solves
instances $(U,\cF_\text{SP},t)$ of 
$k$-\textsc{Set Partitioning} with $\cF_\text{SP}$ down-closed in time 
$\bO((2-\beta)^n|\cF_\text{SP}|\poly(n))$.
Then there exists an algorithm that solves instances $(U,\cF_\text{SC},t)$
of $k$-\textsc{Set Cover} in time
$\bO(((2-\beta)^n+2^k)|\cF_\text{SC}|\poly(n))$.
\end{Lem}
\begin{Proof}
Let $(U,\cF_\text{SC},t)$ be a given input to $k$-\textsc{Set Cover}. 
Compute the down-closure of $\cF_\text{SC}$ to obtain
$\cF_\text{SP}=\down\cF_\text{SC}$. Observe that 
$|\cF_\text{SP}|\leq 2^k|\cF_\text{SC}|$ and that 
this down-closure can be computed in $\bO(2^k|\cF_{\text{SC}}|\poly(n))$ time 
by considering each set in $\cF_\text{SC}$ in turn and listing its up 
to $2^k$ subsets; furthermore, all sets in $\cF_\text{SP}$ have size at 
most $k$. Next, run the assumed algorithm for 
$k$-\textsc{Set Partitioning} on the instance $(U,\cF_\text{SP},t)$ 
and report its output as the output for $k$-\textsc{Set Cover}
on the given input $(U,\cF_\text{SC},t)$.
By down-closure/down-closedness, we have that 
$(U,\cF_\text{SC},t)$ is a YES-instance of $k$-\textsc{Set Cover}
if and only if the constructed $(U,\cF_\text{SP},t)$ is a YES-instance of 
$k$-\textsc{Set Partitioning}. 
\end{Proof} 

Based on Lemma~\ref{lem:covering-to-partitioning}, to prove 
Theorem~\ref{thm:main} it now suffices to solve the 
$k$-\textsc{Set Partitioning} problem for $k\leq \alpha n$ 
in time $\bO^*((2-\beta)^n)$ for constants $\alpha,\beta>0$. 
Indeed, for $k\leq\kappa n$ we have 
$|\cF_\text{SP}|,|\cF_\text{SC}|\leq 2^{H(\kappa)n}$,
so choosing the constant $0<\kappa<\alpha$ small enough enables 
a constant $\epsilon>0$ with 
\begin{equation}
\label{eq:epsilon-choice}
(2-\beta)2^\kappa 2^{H(\kappa)}<2-\epsilon\,.
\end{equation}

We will now reduce to the algorithm in Lemma~\ref{lem:main-three-way}, 
with the parameters $\tau=\frac{1}{1000}>0$ and $\delta=\frac{1}{100000}>0$ 
derived in Section~\ref{sect:broken-randomize}.
Let $\alpha=\tau$ and 
let $(U,\cF_\text{SP},t)$ be an instance of the 
$k$-\textsc{Set Partitioning} problem for $k\leq \alpha n$
given as input.

Let us study an arbitrary but fixed solution $\mathcal{S}$ of the 
instance $(U,\mathcal F_{\mbox{\tiny SP}},t)$. We observe that 
$\mathcal{S}$ can be partitioned into three 
parts $\mathcal{P}_1,\mathcal{P}_2,\mathcal{P}_3$ 
with $a_i=|\!\cup_{S\in \mathcal{P}_i} S|\leq \frac{n}{3}+k$ for 
all $i=1,2,3$. Indeed, this can be realized by assuming the opposite, and 
moving any set from a too large part $\mathcal{P}_i$ 
with $a_i>\frac{n}{3}+k$ to a part $\mathcal{P}_j$ 
with the smallest $a_j$, which must have $a_j< \frac{n}{3}-k/2$ by 
an averaging argument. Hence, any such moving of a set---which is of size 
at most $k$---will make $a_i$ smaller and will not make $a_j$ too large. 
We can repeat such moves until no part is too large and thus obtain
$\mathcal{P}_1,\mathcal{P}_2,\mathcal{P}_3$ from $\mathcal{S}$.

We now reduce to three-way partitioning. 
First, for all nonempty subsets $X\subseteq U$ 
with $|X|\leq (\frac{1}{3}+\tau)n$,
let $C[X]$ be the minimum positive integer $\ell$ for which there exist
$S_1,S_2,\ldots,S_\ell\in\cF_\text{SP}$ with 
$S_1\dcup S_2 \dcup\cdots\dcup S_\ell=X$. 
We can compute $C[X]$ for all $X$ by proceeding from the base 
case $C[\emptyset]=0$ via the following dynamic-programming recurrence
\begin{equation}
\label{eq:dp-c}
C[X]=\min_{\substack{Y\in \cF_\text{SP}\\ Y\subseteq X}} (C[X\setminus Y]+1)\,.
\end{equation}
The computation of \eqref{eq:dp-c} over all $X$ can be implemented in time 
$\bO(2^{H(1/3+\tau)n}|\cF_\text{SP}|\poly(n))\subseteq\bO(2^{H(1/3+\tau)n}2^{H(\tau)n}\poly(n))$.
With our choice of $\tau=\frac{1}{1000}$, this part of 
the reduction runs in $\bO(1.91^n)$ time. 

Second, we construct $t$ set families $\cC_1,\cC_2,\ldots,\cC_t$ with 
\[
\cC_i=\{X\subseteq U:1\leq |X|\leq (1/3+\tau)n,\ C[X]=i\}
\]
for $i=1,2,\ldots,t$. Next, we iterate over all triples $(t_1,t_2,t_3)$ with 
$1\leq t_1\leq t_2 \leq t_3\leq t-2$ and $t_1+t_2+t_3\leq t$.
For each triple $(t_1,t_2,t_3)$, we run the three-way-partitioning 
algorithm from Lemma~\ref{lem:main-three-way} 
with the input $\cF=\cC_{t_1}$, $\cG=\cC_{t_2}$ and $\cH=\cC_{t_3}$. 
Whenever the algorithm of Lemma~\ref{lem:covering-to-partitioning} 
reports a solution, we have also found a solution to instance
$(U,\cF_\text{SP},t)$. Conversely, we know that an arbitrary fixed 
solution $\mathcal{S}$ to $(U,\cF_\text{SP},t)$ can be partitioned into 
three parts $\mathcal{P}_1,\mathcal{P}_2,\mathcal{P}_3$ with
$t_i=C[\cup_{S\in \mathcal{P}_i} S]$ for $i=1,2,3$ as well as
$1\leq t_1\leq t_2\leq t_3\leq t-2$ and $t_1+t_2+t_3\leq t$. 
Since we consider all such three-tuples $(t_1,t_2,t_3)$, 
we will eventually find the correct one. Without loss of generality
we can assume that $t_i<t\leq n$, so there are at most $n^3$ such three-tuples.
Thus, the final part of the reduction runs in $O^*((2-\delta)^n)$ time
and succeeds with probability at least $1-\exp(-\Omega(n))$. 
In particular, we can solve the $k$-\textsc{Set Partitioning} problem 
for $k\leq \alpha n$ in time $\bO^*((2-\beta)^n)$ for $\beta=\delta$.

From $\alpha=\frac{1}{1000}$, $\beta=\frac{1}{100000}$, and 
\eqref{eq:epsilon-choice} it follows we can 
take $\epsilon=\kappa=\frac{1}{10000000}$ in Theorem~\ref{thm:main}, again
stressing that we have optimized neither the analysis nor the construction
for the best possible constants. To mention one way to improve the bounds, we could use faster but more complicated algorithms known to compute $C[X]$, cf.~\cite{BjorklundHKK10}.
This completes the proof of Theorem~\ref{thm:main}.

\section*{Acknowledgements}

Much of our understanding of exponential time problems, the set cover 
conjecture, and their relations to other problems, have been acquired during    
Dagstuhl Seminar 10441 (Exact Complexity of NP-hard Problems), 
Dagstuhl Seminar 13331 (Exponential Algorithms: Algorithms and Complexity Beyond Polynomial Time), 
Dagstuhl Seminar 16451 (Structure and Hardness in P), and
a one-semester program at the Simons Institute for the Theory of Computing 
(Fine-grained Complexity and Algorithm Design).
We would also like to thank the organizers and participants of 
the 2023 Workshop on Algebra and Computation at Chalmers University.

AB is supported by the VILLUM Foundation, Grant 16582.


\bibliographystyle{abbrv}
\bibliography{paper.bib}


\clearpage

\appendix

\begin{center}
\textsc{\large Appendix}
\end{center}

\section{Proof of Lemma~\ref{lem:three-way} for the complex numbers}

\label{sect:complex}

This appendix proves Lemma~\ref{lem:three-way} in the case when $\F$
is the field $\C$ of complex numbers. In particular, it suffices to present
a version of the broken algorithm (recall Sect.~\ref{sect:broken-algorithm}) 
suitable for this case. 
Let us write $\Q$ for the field of rational numbers; for purposes of
computation in what follows, we assume that elements of $\Q$ are represented 
in binary as a numerator-denominator pair of coprime integers in explicit 
binary positional representation.

Let $\rho>0$ be a constant whose precise value will be fixed later. 
Let $Q\in\Q^{7\times 7\times 7}\subseteq\C^{7\times 7\times 7}$ be 
the tensor \eqref{eq:q-tensor}.
Under the asymptotic rank conjecture over $\C$, there
exist positive integer constants $d$ and $7^d\leq r\leq 7^{(1+\rho)d}$ 
as well as three matrices $\alpha,\beta,\gamma$ of polynomial indeterminates, 
each of shape $7^d\times r$, such that the system of cubic polynomial 
equations
\begin{equation}
\label{eq:rank-eq}
\sum_{\ell\in[r]}\alpha_{i,\ell}\beta_{j,\ell}\gamma_{k,\ell}-
(Q^{\otimes d})_{i,j,k}=0\qquad\text{for all $i,j,k\in [7^d]$}
\end{equation}
has a solution over $\C$.

Let $I\subseteq \C[\alpha_{i,\ell},\beta_{i,\ell},\gamma_{i,\ell}:i\in[7^d],\,\ell\in[r]]$ be the polynomial ideal 
generated by the left-hand-side polynomials of the system \eqref{eq:rank-eq}.
By Hilbert's Nullstellensatz~(see e.g.~\cite[Chap.~4, \S1]{CoxLO2015}), we have 
$1\notin I$ since \eqref{eq:rank-eq} has a solution over the complex numbers.
Let $G\subseteq \C[\alpha_{i,\ell},\beta_{i,\ell},\gamma_{i,\ell}:i\in[7^d],\,\ell\in[r]]$ be a reduced Gr\"obner basis 
for $I$. In particular, using, for example, Buchberger's 
algorithm~(see e.g.~\cite[Chap.~2, \S7]{CoxLO2015}), we can compute $G$ from 
the generators \eqref{eq:rank-eq} in constant time; indeed, recall that 
$\rho$, $d$, and $r$ are constants and $Q$ is a constant-size tensor. 
Furthermore, by the structure of 
Buchberger's algorithm, the elements of $G$
are {\em rational-coefficient} polynomials. That is, we have
$G\subseteq\Q[\alpha_{i,\ell},\beta_{i,\ell},\gamma_{i,\ell}:i\in[7^d],\,\ell\in[r]]$ and $I=\bra G\ket$.

Let $n$ be a nonnegative integer and let $\cF,\cG,\cH\subseteq 2^{[n]}$
be given as input. Let $\epsilon$ be a further polynomial indeterminate. 
We will work in the rational-coefficient polynomial quotient ring 
\begin{equation}
\label{eq:complex-case-ring}
\Ri=\Q[\epsilon,\alpha_{i,\ell},\beta_{i,\ell},\gamma_{i,\ell}:i\in[7^d],\,\ell\in[r]]/I\,.
\end{equation}
In particular, addition in $\Ri$ reduces to addition of rational-coefficient
multivariate polynomials, and multiplication in $\Ri$ reduces to multiplication
of rational-coefficient multivariate polynomials followed by multivariate 
polynomial remainder (see e.g.~\cite[Chap.~2, \S3]{CoxLO2015}) with respect 
to the Gr\"obner basis $G$ and under an arbitrary but fixed monomial 
ordering (see e.g.~\cite[Chap.~2, \S2]{CoxLO2015}). 

To simplify our analysis
of the algorithm to follow---to avoid analysing intermediate 
rational-coefficient arithmetic in particular, and at the extra cost of 
a worse polynomial-in-$n$ factor to the running time---%
in our algorithm we perform arithmetic in $\Ri$ so that we do {\em not} reduce 
multivariate polynomial representatives of $I$-equivalence classes by 
multivariate polynomial remaindering 
with respect to $G$ until at the very end, where we perform one remaindering
with respect to $G$ to obtain the output value $\Gamma$ in $\Ri$.
In particular, as we will see, the algorithm will work over 
the {\em integer} polynomial ring
$\Z[\epsilon,\alpha_{i,\ell},\beta_{i,\ell},\gamma_{i,\ell}:i\in[7^d],\,\ell\in[r]]$ until it is ready to produce its output, at which point we will 
extend to rationals and run remaindering with respect to the 
(rational-coefficient) polynomials in $G$, requiring an analysis of the 
bit-complexity of the rational coefficients during this final remaindering step.

To set up the algorithm, 
let us write $\alpha,\beta,\gamma\in\Ri^{7^d\times r}$ for the $7^d\times r$
matrices with the $(i,\ell)$-entry equaling to the corresponding polynomial 
indeterminates $\alpha_{i,\ell},\beta_{i,\ell},\gamma_{i,\ell}$
of $\Ri$, respectively, for each $i\in[7^d]$ and $\ell\in [r]$. 
By the construction of $I\not\ni 1$ and \eqref{eq:complex-case-ring}, 
over $\Ri$ we thus have 
\begin{equation}
\label{eq:k-alpha-beta-gamma}
\K(\alpha,\beta,\gamma)=Q^{\otimes d}\,.
\end{equation}
We also recall that $\epsilon$ is a polynomial indeterminate in $\Ri$ and
let us write $A_\epsilon,B_\epsilon,C_\epsilon\in\Ri^{2\times 2}$
for the three matrices in~\eqref{eq:three-way-partitioning-border-2}. 

We are now ready to describe a version of the broken algorithm 
for the complex case. However, we stress that all arithmetic in the algorithm 
takes place over multivariate polynomials, first with {\em integer} coefficients
and then, in the final reduction step, with {\em rational} coefficients. 
We recall the underline-notation from Sect.~\ref{sect:broken-algorithm} and 
the correspondence \eqref{eq:x-underline} in particular.

We may again assume that $n=p+3dq$ for two nonnegative integers $p$ and $q$. 
First, the algorithm removes all
the bad sets in each of the families $\cF,\cG,\cH$ given as input; let us
write $\underline{\cF},\underline{\cG},\underline{\cH}$ for the set families
so obtained. Second, the algorithm constructs the zero-one-integer-valued 
vectors $\underline{f},\underline{g},\underline{h}\in\Z^{2^p7^{dq}}$ defined
for all $\underline{x},\underline{y},\underline{z}\in [2^p7^{dq}]$ by
\begin{equation}
\label{eq:complex-fgh-underline}
\underline{f}_{\underline{x}}=\iv{X\in\underline{\cF}}\,,\quad
\underline{g}_{\underline{y}}=\iv{Y\in\underline{\cG}}\,,\quad
\underline{h}_{\underline{z}}=\iv{Z\in\underline{\cH}}\,.
\end{equation}
Third, using Yates's algorithm (Sect.~\ref{sect:yates}) over the ring
$\Z[\epsilon,\alpha_{i,\ell},\beta_{i,\ell},\gamma_{i,\ell}:i\in[7^d],\,\ell\in[r]]$
in each case, the algorithm computes the three matrix-vector products
\begin{equation}
\label{eq:complex-broken-yates}
\begin{split}
\tilde f_{\epsilon,\alpha}&\YL ((A_\epsilon^\top)^{\otimes p}\otimes (\alpha^\top)^{\otimes q})\underline f\,,\\
\tilde g_{\epsilon,\beta}&\YL ((B_\epsilon^\top)^{\otimes p}\otimes (\beta^\top)^{\otimes q})\underline g\,,\\
\tilde h_{\epsilon,\gamma}&\YL ((C_\epsilon^\top)^{\otimes p}\otimes (\gamma^\top)^{\otimes q})\underline h\,.
\end{split}
\end{equation}
Fourth, the algorithm computes the 
$\Z[\alpha_{i,\ell},\beta_{i,\ell},\gamma_{i,\ell}:i\in[7^d],\,\ell\in[r]]$-polynomial
\begin{equation}
\label{eq:complex-broken-output}
\Gamma\leftarrow\{\epsilon^p\}(\tilde f_{\epsilon,\alpha}^\top(\tilde g_{\epsilon,\beta}\odot\tilde  h_{\epsilon,\gamma}))\,.
\end{equation}
Finally, the algorithm reduces $\Gamma$ by multivariate polynomial 
remaindering with respect to the Gr\"obner basis $G$ to obtain the
reduced $\Gamma\in\Ri$; recall~\eqref{eq:complex-case-ring}.
The algorithm outputs that a good three-way partition exists if
$\Gamma\neq 0$ over $\Ri$; otherwise the algorithm outputs that no good 
three-way partition exits. This completes the description of the algorithm.
We observe in particular that only the final reduction step of the algorithm 
runs rational-coefficient polynomial arithmetic, all other polynomials 
manipulated by the algorithm have integer coefficients.

To establish correctness, first observe that since $\Z\subseteq\Q$ we
can analyse the entire algorithm as if it operated over $\Ri$. Then, 
proceeding as in the proof of Lemma~\ref{lem:broken-on-good}, from 
\eqref{eq:three-way-partitioning-border-2}, \eqref{eq:k-alpha-beta-gamma},
\eqref{eq:complex-fgh-underline}, and \eqref{eq:t-good-partitioning}
we conclude that 
for the tensor $T=P^{\otimes p}\otimes Q^{\otimes dq}$ over $\Ri$
we have
\[
\Gamma
=T[\underline{f},\underline{g},\underline{h}]
=\sum_{\underline{x},\underline{y},\underline{z}\in[2^p7^{dq}]}
T_{\underline{x},\underline{y},\underline{z}}
\underline{f}_{\underline{x}}
\underline{g}_{\underline{y}}
\underline{h}_{\underline{z}}
=\sum_{\substack{%
(X,Y,Z)\in\underline{\cF}\times \underline{\cG}\times \underline{\cH}\\
X\dcup Y\dcup Z=[n]}} 1\,.
\]
That is, $\Gamma\in\Q\subseteq\Ri$ counts the number of good three-way
partitions in the input $\cF,\cG,\cH$. The algorithm is thus correct
on good three-way partitions. 

Proceeding as in Sect.~\ref{sect:broken-algorithm} and 
Sect.~\ref{sect:broken-randomize} to analyse the running time and run 
independent repeats of the algorithm on randomly permuted inputs, we obtain 
Lemma~\ref{lem:three-way} for the complex case subject to the assumption
that no arithmetic operation on $\Z$-polynomial 
(and in the final remaindering step, $\Q$-polynomial) operands executed 
by the algorithm takes more than $\poly(n)$ time for a fixed polynomial 
that depends only on $n$ (as well as on the constants $d$ and $r$---the 
dependence on $d$ and $r$ is {\em not} polynomial, however). 

To justify the assumption, let us first study the degree and the number of
monomials involved in the polynomial operands, without yet paying attention
to the coefficient complexity. We  
recall from \eqref{eq:complex-case-ring} that $\Ri$ is a polynomial quotient
ring with rational coefficients over a constant number 
(depending only on $d$ and $r$) of polynomial indeterminates. Also, the
Gr\"oebner basis $G$ has constant (depending only on $d$ and $r$) size. 
The elements of each of the matrices 
$\alpha,\beta,\gamma,A_\epsilon,B_\epsilon,C_\epsilon$ as well as 
the initial vectors $\underline{f},\underline{g},\underline{h}$
are polynomials of degree at most $1$.
Since $p,q\leq n$, it is thus immediate 
from \eqref{eq:complex-fgh-underline},
\eqref{eq:complex-broken-yates}, and \eqref{eq:complex-broken-output}
that all intermediate results are polynomials
of degree at most $3n$ in a constant $D=3\cdot 7^{d}r+1$ 
number of indeterminates; in particular, each such polynomial has at 
most $(3n+1)^D$ monomials with a nonzero coefficient.
Thus, we are done if we can conclude that every arithmetic operation
on coefficients takes at most $\poly(n)$ time.

For the part of the algorithm working with integer coefficients, it is 
immediate that every integer operand is at most $\poly(n)$ bits. 
Indeed, we start with one-bit integers in \eqref{eq:complex-fgh-underline},
and thus \eqref{eq:complex-broken-yates} proceeds to at most 
$\bO(n)$-bit integer coefficients for each monomial in each polynomial 
since the monomial-entried matrices 
$\alpha,\beta,\gamma,A_\epsilon,B_\epsilon,C_\epsilon$ have 
$\bO(1)$-bit integer coefficients for each monomial. Thus, an $\bO(n)$-bit bound
for the integer coefficient of each monomial persists 
in the course of \eqref{eq:complex-broken-output}. 

It remains to conclude that arithmetic on rational coefficients in the 
final remaindering step with respect to $G$ (which consists 
of $\bO(1)$ rational-coefficient polynomials, each of degree $\bO(1)$, over 
$\bO(1)$ indeterminates, and with $\bO(1)$-bit rational coefficients). The
polynomial $\Gamma$ to be remaindered with respect to $G$ is over 
$\bO(1)$ indeterminates and has at most $\poly(n)$ degree and at 
most $\poly(n)$ monomials with integer coefficients of at most $\poly(n)$ bits. 

By the structure of multivariate remaindering 
(see~\cite[Chap.~2, \S3, Theorem~3]{CoxLO2015}), it suffices to prove the
following claim. Suppose that $T$ is an array of 
$N\leq \poly(n)$ rational numbers that is 
initially populated with {\em integers} with at most $\poly(n)$ bits each.
Let $R$ be an array of $L=\bO(1)$ rational numbers with $\bO(1)$ bits each.
Suppose we run a sequence of at most $\poly(n)$ arbitrary operations of 
either of the following two types on the array: 
\begin{enumerate}
\item[(i)] for some choice of $i,j,k\in [N]$, let $T_k\leftarrow T_i+T_j$; or 
\item[(ii)] for some choice of $i,j\in [N]$ and $\ell\in [L]$, 
let $T_j\leftarrow R_\ell\cdot T_i$.
\end{enumerate}
The claim is that the sequence of operations takes in total 
at most $\poly(n)$ time.

To see that the claim suffices to control the multivariate-polynomial 
remaindering of $\Gamma$ 
with respect to $G$, observe that $T$ can be assumed to initially contain
all the coefficients of the monomials of the polynomial $\Gamma$, as well 
as all the working space needed by the remaindering algorithm. 
The array $R$ contains working constants
(such as the constant $-1$ to enable subtraction using one (ii)-operation 
followed by an (i)-operation) as well as the ratios of all pairs of 
coefficients (including the coefficient $1$ to obtain the coefficients 
themselves as well as their inverses among the ratios) of monomials of 
polynomials in $G$; there are $\bO(1)$ such ratios and 
each uses $\bO(1)$ bits for its rational representation. 
Indeed, the remaindering algorithm 
(see~\cite[Chap.~2, \S3, Theorem~3]{CoxLO2015})
operates by coefficient-arithmetic 
operations of the form $T_k\leftarrow T_i\pm T_j\sigma^{-1}$,
$T_k\leftarrow T_i\pm T_j\sigma^{-1}\tau$, and $T_k\leftarrow T_i\pm T_j$,
where $\sigma,\tau$ are coefficients of monomials in polynomials in $G$;
each of these operations can be implemented with $\bO(1)$ operations
of types (i) and (ii).

To establish the claim, let us study the rational numbers in the $R$ array. 
Let $p_1,p_2,\ldots,p_c$ be the distinct prime numbers that occur in 
at least one of the denominators of the numbers in $R$. 
We have $p_j\leq \bO(1)$ for all $1\leq j\leq c=\bO(1)$.
In particular, each rational number in the array $R$ has the form
\begin{equation}
\label{eq:r-rep}
\frac{r}{p_1^{d_1}p_2^{d_2}\cdots p_c^{d_c}}
\end{equation}
for some integer $r\leq \bO(1)$ and nonnegative integers $d_j\leq \bO(1)$ for
$1\leq j\leq c$.

Let us now study the operations of types (i) and (ii) on the array $T$.
For two integers $x$ and $y$, for brevity let us write $x\vee y$ for 
the maximum of $x$ and $y$, and write $x\overline{\vee} y$ for $0\vee (x-y)$.
We will work in a numerator-denominator representation motivated 
by \eqref{eq:r-rep}. Let $t$ and $t'$ be integers and let 
$e_1,e_2,\ldots,e_c$ and $e_1',e_2',\ldots,e_c'$ be nonnegative integers.
Over the rationals, we have
\begin{equation}
\label{eq:r-rep-sum}
\frac{t}{p_1^{e_1}p_2^{e_2}\cdots p_c^{e_c}}
+
\frac{t'}{p_1^{e_1'}p_2^{e_2'}\cdots p_c^{e_c'}}
=
\frac{%
tp_1^{e_1'\overline{\vee} e_1}p_2^{e_2'\overline{\vee} e_2}\cdots p_1^{e_c'\overline{\vee} e_c}\ 
+\ 
t'p_1^{e_1\overline{\vee} e_1'}p_2^{e_2\overline{\vee} e_2'}\cdots p_1^{e_c\overline{\vee} e_c'}
}{p_1^{e_1\vee e_1'}p_2^{e_2\vee e_2'}\cdots p_c^{e_c\vee e_c'}}\,.
\end{equation}
Similarly, we have 
\begin{equation}
\label{eq:r-rep-prod}
\frac{r}{p_1^{d_1}p_2^{d_2}\cdots p_c^{d_c}}
\cdot
\frac{t}{p_1^{e_1}p_2^{e_2}\cdots p_c^{e_c}}
=
\frac{rt}{p_1^{d_1+e_1}p_2^{d_2+e_2}\cdots p_c^{d_c+e_c}}\,.
\end{equation}
In particular, from \eqref{eq:r-rep-sum} and \eqref{eq:r-rep-prod} we 
immediately observe that we can assume that
the rational form $\frac{t}{p_1^{e_1}p_2^{e_2}\cdots p_c^{e_c}}$ for 
an integer $t$ and nonnegative integers $e_1,e_2,\ldots,e_c$ 
persists in the entries of the array $T$ when operations of 
types (i) and (ii) are executed on the array.

To parameterize the analysis of steps (i) and (ii), let us write 
$B_s$ for the longest bit-length of a rational numerator in the array $T$
at step $s=0,1,\ldots$ of the sequence of operations, and
$E_s$ for the maximum exponent of a prime in the prime factorization of 
any of the denominators of the rationals in $T$.
Observe in particular that initially 
\begin{equation}
\label{eq:start-bound}
B_0\leq\poly(n)
\qquad\text{and}\qquad
E_0=0\,.
\end{equation}
For $s=1,2,\ldots$, from~\eqref{eq:r-rep-sum} 
as well as $p_j\leq \bO(1)$ for all $1\leq j\leq c=\bO(1)$ we have
\begin{equation}
\label{eq:type-i-bound}
B_s\leq B_{s-1}+\bO(1)\cdot E_{s-1}+\bO(1)
\qquad\text{and}\qquad
E_s=E_{s-1}
\end{equation}
when operation $s$ in the sequence of operations has type (i) and
 \eqref{eq:r-rep-prod} we have
\begin{equation}
\label{eq:type-ii-bound}
B_s\leq B_{s-1}+\bO(1)
\qquad\text{and}\qquad
E_s\leq E_{s-1}+\bO(1)
\end{equation}
when operation $s$ in the sequence of operations has type (ii). 
Chaining the inequalities \eqref{eq:type-i-bound} and
\eqref{eq:type-ii-bound} with the base case \eqref{eq:start-bound}, 
we have that 
$B_s,E_s\leq\poly(n)$ holds for $s\leq\poly(n)$. Thus, since the denominators
of rationals in $T$ involve only the primes $p_1,p_2,\ldots,p_c\leq \bO(1)$, 
we have that 
for $s\leq\poly(n)$ each entry of $T$ is representable using at most 
$B_s+\bO(1)\cdot E_s+\bO(1) \leq \poly(n)$
bits. This completes the proof of the claim and thus the proof of 
Lemma~\ref{lem:three-way} when $\F=\C$.


\end{document}